\documentclass[a4paper]{elsarticle}
\usepackage{amsmath,amssymb,amsfonts}
\usepackage{calrsfs}
\usepackage{algorithm}
\usepackage{algpseudocode}
\usepackage{lineno,hyperref}
\usepackage[usenames]{xcolor}
\usepackage{xspace}

\renewcommand{\Re}{\operatorname{Re}}
\renewcommand{\Im}{\operatorname{Im}}

\newcommand{\ie}{i.e.\@\xspace}

\modulolinenumbers[5]

\newcommand\edited[1]{#1}

\journal{Applied and Computational Harmonic Analysis}

\def\inv{{^{-1}}}

\def\RR{{\mathbb R}}

\def\EE{{\mathbb E}}

\def\Ex#1{{\EE\left\{#1\right\}}}

\def\CC{\mathbb C}

\def\NN{\mathbb N}
\def\Id{{\mathsf I}}
\def\Tr#1{{\mathsf{Tr}\!\left(#1\right)}}

\def\defeq{\stackrel{\Delta}{=}}

\DeclareMathAlphabet{\mathpzc}{OT1}{pzc}{m}{it}

\def\cB{{\mathcal B}}

\def\cD{{\mathcal D}}

\def\cCN{{\mathcal{C\!\!N}}}

\def\cL{{\mathcal L}}

\def\cM{{\mathcal M}}

\def\cN{{\mathcal N}}

\def\cS{{\mathcal S}}

\def\cW{{\mathcal W}}

\def\Lone{L^1(\RR )}
\def\Ltwo{L^2(\RR )}

\def\bA{{\mathbf A}}
\def\bb{{\mathbf b}}
\def\bB{{\mathbf B}}
\def\bC{{\mathbf C}}
\def\tbC{\widetilde{\mathbf C}}
\def\bD{{\mathbf D}}
\def\obD{\overline{\mathbf D}}
\def\be{{\mathbf e}}

\def\bG{{\mathbf G}}
\def\tbG{\widetilde{\bG}}

\def\bm{{\mathbf m}}
\def\tbm{\widetilde{\bm}}

\def\bR{{\mathbf R}}
\def\bRpo{{\mathbf R}_\mathsf{po}}
\def\tbRpo{\widetilde{\mathbf R}_\mathsf{po}}

\def\bw{{\mathbf w}}
\def\tbw{\widetilde{{\mathbf w}}}
\def\bW{{\mathbf W}}
\def\ubW{\underline{{\mathbf W}}}
\def\tbW{\widetilde{{\mathbf W}}}
\def\bx{{\mathbf x}}
\def\bX{{\mathbf X}}
\def\by{{\mathbf y}}
\def\bY{{\mathbf Y}}
\def\bz{{\mathbf z}}
\def\bZ{{\mathbf Z}}

\def\beps{\boldsymbol{\varepsilon}}
\def\bGamma{\boldsymbol{\Gamma}}
\def\bGammapr{\bGamma_\mathsf{pr}}
\def\bGammapo{\bGamma_\mathsf{po}}
\def\obGammapr{\overline{\bGamma}_\mathsf{pr}}
\def\obGammapo{\overline{\bGamma}_\mathsf{po}}
\def\tbGammapo{\widetilde{\bGamma}_\mathsf{po}}
\def\tbGamma{\widetilde{\boldsymbol{\Gamma}}}

\def\bPsi{\boldsymbol{\Psi}}

\def\bmu{\boldsymbol{\mu}}
\def\tbmu{\widetilde{\boldsymbol{\mu}}}

\def\tbmupo{\widetilde{\boldsymbol{\mu}}_\mathsf{po}}

\def\bvtheta{\boldsymbol{\vartheta}}

\def\tbvtheta{\widetilde{\boldsymbol{\vartheta}}}
\def\bTheta{\boldsymbol{\Theta}}

\newtheorem{proposition}{Proposition}
\newtheorem{remark}{Remark}

\begin{document}

\begin{frontmatter}

\title{Synthesis-based time-scale transforms for non-stationary signals}

\author{Adrien Meynard\fnref{Afootnote}}
\address{ENS de Lyon, CNRS, Laboratoire de physique, Lyon, France}
\fntext[Afootnote]{since September 2021, formerly at the Department of Mathematics, Duke University, Durham, NC, USA}
\ead{adrien.meynard@ens-lyon.fr}

\author{Bruno Torr\'esani}
\address{Aix-Marseille Univ, CNRS, I2M, Marseille, France}
\ead{bruno.torresani@univ-amu.fr}




\begin{abstract}
This paper deals with the modeling of non-stationary signals, from the point of view of signal synthesis. A class of random, non-stationary signals, generated by synthesis from a random time-scale representation, is introduced and studied. Non-stationarity is implemented in the time-scale representation through a prior distribution which models the action of time warping on a stationary signal. A main originality of the approach is that models directly a time-scale representation from which signals can be synthesized, instead of post-processing a pre-computed time-scale transform.

A \textit{maximum a posteriori} estimator is proposed for the time warping parameters and the power spectrum of an underlying stationary signal, together with an iterative algorithm, called JEFAS-S, for the estimation, based upon the Expectation Maximization approach.

Numerical results show the ability of JEFAS-S to estimate accurately time warping and power spectrum. This is in particular true when time warping involves fast variations, where a similar approach called JEFAS, proposed earlier, fails. In addition, as a by-product, the approach is able to yield extremely sharp time-scale representations, also in the case of fast varying non-stationarity, where standard approaches such as synchrosqueezing fail.
\end{abstract}


\end{frontmatter}


\section{Introduction}

In the statistical signal processing or time series analysis literature, the spectral analysis problem aims at estimating the spectral content of a stationary signal or time series, called power spectrum (in other words, the distribution of power as a function of frequency) from a discrete, finite-length sample. The existence of such a power spectrum is guaranteed by the Wiener-Khinchin theorem. In addition, under suitable assumptions, random signals may also be represented in terms of filtering of white noise, where the filter's frequency response is defined by the square root of the power spectrum. This representation is known as Cram\' er's representation, we refer to~\cite{Koopmans1995spectral} for a thorough presentation. In this context,  spectral analysis may be addressed by parametric models (for example AR or ARMA), or nonparametric approaches (see e.g.~\cite{Koopmans1995spectral,Stoica1997introduction}). In the latter case, one generally resorts to Fourier-based techniques such as periodograms, involving windowing and time averaging.

However, many signals are not stationary, and these tools do not apply anymore. The extension of spectral analysis to non-stationary situations has been considered by several authors (see e.g.~\cite{Priestley1988nonlinear,Dahlhaus2000likelihood}), who proposed generalizations adapted to specific classes of non-stationary signals.

Among possible approaches, time-frequency and time-scale analysis~\cite{Flandrin2018explorations,Boashash2015time,Carmona1998practical} has received considerable attention during the last decades, as it allows one to step away from traditional stationary signal models expressed as linear combinations of sine waves.
Linear time-frequency or time-scale transforms have been widely used in applications, to provide visualizations from which non-stationary features can be estimated. They also benefit from a strong, well established mathematical support (see e.g.~\cite{Groechenig2001foundations}). A main aspect of linear transform is the existence, under suitable assumptions, of inverse mappings, which express signals as linear combinations of time-frequency localized waveforms. These transforms are known to be limited by uncertainty principles: even though the classical Heisenberg inequality does not explicitly involve  a joint time-frequency domain, the short-time Fourier transform (to quote an example) can be viewed as the output of a filter bank, and the impulse responses of the filters are affected by Heisenberg's inequality. Super-resolution techniques such as reassignment or synchrosqueezing~\cite{Auger2013time} appear as mere post-processing of one or several linear transforms affected by uncertainty, which attempt to correct the loss of time-frequency resolution by suitable non-linear modifications. An important asset of synchrosqueezing is the preservation of invertibility: reconstruction is possible from synchrosqueezed wavelet or short time Fourier transforms.

Amplitude and Frequency modulated (AM-FM) signal models have received considerable attention as well, often in different contexts.  Originally defined in terms of the Hilbert transform (see~\cite{Picinbono1998remarks}), instantaneous amplitude and frequency can also be estimated using such tools. The problem was later on addressed through time-frequency transforms~\cite{Delprat1993asymptotic,Carmona1998practical} and super-resolution techniques~\cite{Auger2013time}. More generally, models involving sums of AM-FM signals have also been thoroughly studied in various contexts, including speech modeling~\cite{McAulay1986speech} and music processing~\cite{Depalle1993analysis}, and more recently for physiological signal analysis~\cite{Wu2013instantaneous} and~\cite{Chui2016signal}. These models express signals as linear combinations of amplitude and frequency modulated ``modes'', to be identified. Again there is a vast literature on the subject which we will not review here, let us just mention empirical mode decomposition methods~\cite{Huang1998empirical}, iterative filtering~\cite{Cicone2016adaptive,Cicone2020iterative} or the frequency extraction approach of~\cite{Chui2016signal}, which have enjoyed significant interest recently.

\begin{figure}
\centering
\includegraphics[scale=.4]{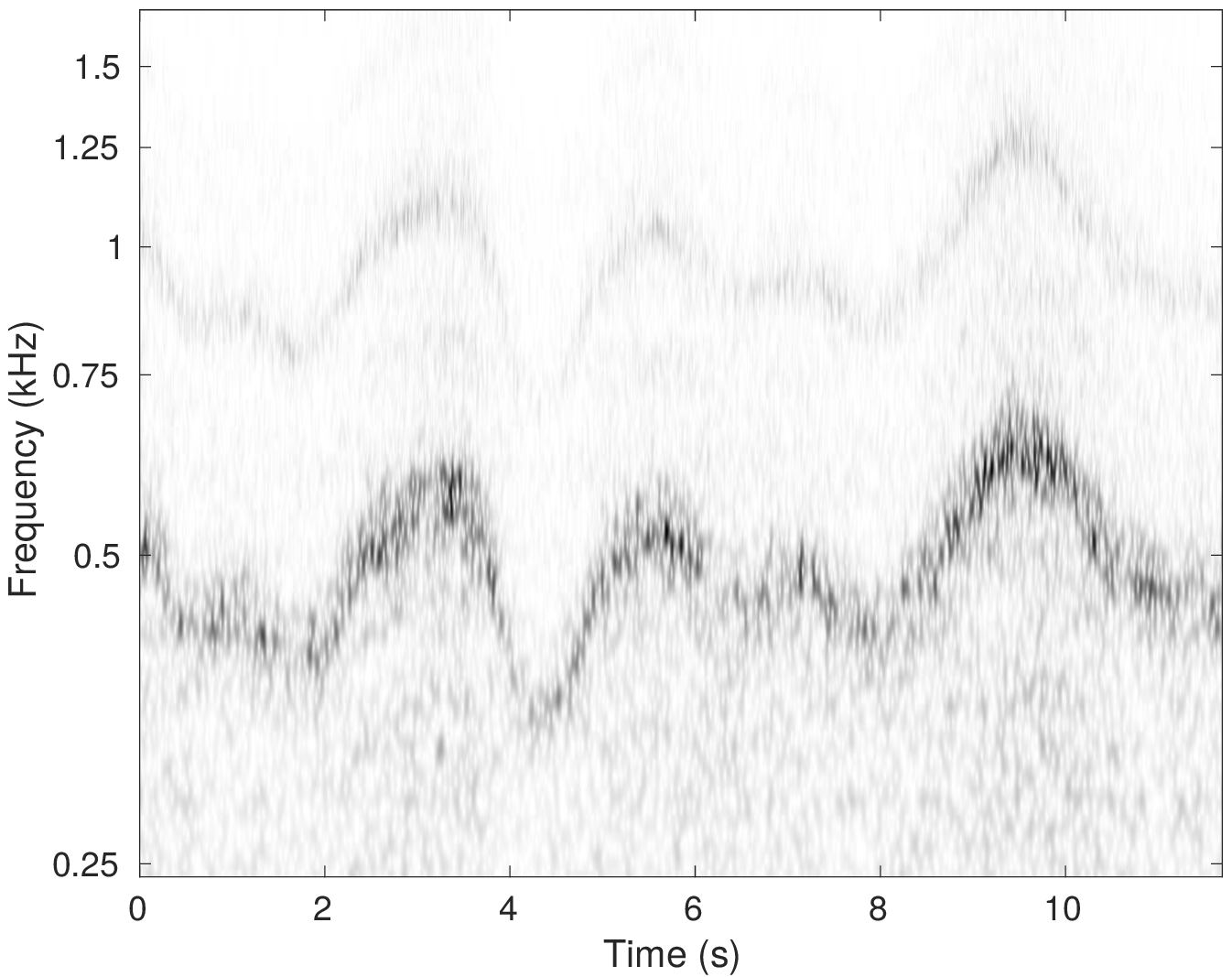}
\caption{Scalogram of a 10-second-long wind recording.}
\label{fi:wind}
\end{figure}

However, not every non-stationary signal can be described as AM-FM or a sum of AM-FM components. Many physically relevant signals indeed involve FM-like non-stationarity, but the latter is combined with a time varying spectrum broader than standard AM-FM signals. An example can be found in Figure~\ref{fi:wind}, which represents the so-called scalogram (squared modulus of wavelet transform) of a 10-second-long wind recording. Besides the larger local bandwidth already mentioned, the time variations of the local spectrum are quite often poorly accounted by amplitude and frequency modulation, and time warping (i.e. time dependent clock change) is often more physically realistic. \edited{For example, the sound of a accelerating car engine is suitably described by time dependent clock change, as well as stationary signals deformed by Doppler effect. One may also argue that time warping is a convenient model to describe non-stationary ECG signals, which originate from heart beats whose rhythm vary as a function of time.}
To the best of our knowledge, the time warping paradigm has been first studied in depth by Clerc and Mallat in~\cite{Clerc2003estimating}, who developed estimators based upon small scale limite of wavelet transform. It has been revisited more recently in~\cite{Omer2017time,Meynard2018spectral}, in a context of Gaussian random signals, thus avoiding the small scale limit. In~\cite{Meynard2018spectral}, an approximate maximum likelihood approach called JEFAS for the joint estimation of time warping and power spectrum has been developed and studied. JEFAS, which assumes slow variations of the amplitude modulation and time warping function, is essentially a post-processing of a time-scale (wavelet) transform.

In this paper, we develop and study an alternative approach, that may be called a \textit{synthesis approach}, inspired by a model developed by Turner \& Sahani~\cite{Turner2014time}, which interprets time-frequency transform in terms of probabilistic inference. The Turner-Sahani was applied to various time-frequency problems. Our model expresses non-stationary signals as the result of a time-frequency or time-scale synthesis from a random time-frequency representation, governed by a prescribed prior distribution, which involves the time warping model. We study the corresponding model, and compute the posterior distribution which is used to construct an estimator for the model parameters. This approach, called JEFAS-S (where the ``S'' stands for ``synthesis''), also outputs an adapted time-scale representation from which signal synthesis is possible. Remarkably enough, JEFAS-S is able to handle time warping models in situations where the time warping function can have much faster variations. In addition, JEFAS-S can also yield time-frequency representations which are much sharper than conventional transforms, and compare well to synchrosqueezed or reassigned transforms, while allowing signal synthesis.

Some of the results presented in this paper were already sketched in~\cite{Meynard2020time}; we refer also to~\cite{Meynard2019stationnarites}, where the AM-TW model and variants were analyzed in details.

This paper is organized as follows. Background material is given in Section~\ref{se:ModelsTools}, including the \edited{\emph{amplitude modulation and time warping} (AM-TW)} model, elements of time-frequency and time-scale analysis, and an account of the JEFAS approach. The main contributions are described in Section~\ref{se:JEFAS-S}, which in particular develops the statistical model, the estimation procedure and corresponding algorithm. Numerical results and discussion are given in Section~\ref{se:NumRes}, and Section~\ref{se:conclu} is devoted to conclusions. Some more technical aspects are developed in the Appendix.

\section{Models and time-frequency tools}
\label{se:ModelsTools}
\subsection{The time warping model}
\edited{Amplitude and frequency modulations are standard tools in signal processing. However,} amplitude and frequency modulated (AM-FM) signals
\[
y(t) = a(t)\cos(\gamma(t))
\]
may also be understood as amplitude modulated and time warped (AM-TW) signals, \edited{time warping being defined as a time-varying clock change.} More precisely
\begin{equation}
\label{fo:time.warping}
y = \cM_\alpha\cD_\gamma x\ ,
\end{equation}
where $\cM_\alpha$ is the amplitude modulation operator (pointwise multiplication with the positive valued function $\alpha$), and $\cD_\gamma$ is the time warping operator (composition with the monotone function $\gamma$, assumed to be differentiable and strictly increasing), and $a=\alpha\sqrt{\gamma'}$:
\begin{equation}
\cM_\alpha x(t) = \alpha(t) x(t)\ ,\qquad \cD_\gamma x(t) = \sqrt{\gamma'(t)} x(\gamma(t))\ .
\end{equation}
Locally harmonic signals of the form $y(t) = \alpha(t)\sum_k \cos(k\gamma(t))$ may  also be described likewise. We stress that time warping is often more physically relevant than frequency modulation, \edited{examples are provided by car engine sounds during acceleration, Doppler effect (which is generated by speed difference), or ECG signals during heart rate variations.}

The AM-TW model was extended to the setting of random signals in~\cite{Clerc2003estimating}, in the form
\begin{equation}
\label{fo:time.warping.rand}
Y = \cM_\alpha\cD_\gamma X\ ,
\end{equation}
here $X$ is a second order wide sense stationary random process, with zero mean, and absolutely continuous spectral measure. The corresponding power spectrum will be denoted by $\cS_X$. In analogy with the spectral estimation problem for a stationary random processes $X$ (estimate the power spectrum $\cS_X$ from one or several realizations of the process), spectral analysis in this context may be described as the problem of joint estimation of $\alpha$, $\gamma$ and the underlying power spectrum $\cS_X$. The problem has been addressed by various authors (see e.g.~\cite{Priestley1988nonlinear,Dahlhaus2000likelihood}), for different signal models. Since our focus is on AM-TW models, we won't discuss these generalizations further, and limit to time-frequency and time-scale transforms which provide well suited representations for AM-TW models.

\subsection{Time-frequency and wavelets}
Let us first consider the continuous time setting. For the sake of simplicity we limit ourselves to time-scale representation, i.e. wavelet analysis and synthesis, similar developments can be made for more general linear time-frequency transforms. We denote by $\psi\in\Lone\cap\Ltwo$ the analysis wavelet, and by $\psi_s$ the scaled wavelets defined by
\[
\psi_s (t) = q^{-s/2} \psi\left(q^{-s} t\right)\ ,\quad s\in\RR\ ,
\]
for some reference scale $q > 1$. The corresponding wavelet transform associates with every $x\in\Ltwo$ the sequence of band-pass filtered signals $\cW_x(s,\cdot)$ defined by the convolution product
\[
\cW_x(s,\cdot) = x * \widetilde{\psi_s}\ ,
\]
where $\widetilde{\psi_s}(t) = \overline{\psi_s}(-t)$.
In the continuous time setting, it is well known that such a transform is invertible provided $\psi$ satisfies a suitable admissibility condition~\cite{Grossmann1984decomposition}. If the admissibility condition is fulfilled, and after suitable normalization of the wavelet $\psi$, this yields the inversion formula
\begin{equation}
\label{fo:wavelet.synthesis}
x = \int_{-\infty}^\infty \psi_s*\cW_x(s,\cdot)\,\frac{ds}s \ .
\end{equation}
As a consequence of the admissibility condition, the wavelet $\psi$ must be zero-mean, i.e. an oscillatory function. Classical choices include the Morlet wavelet (which has to be modified to fulfill the admissibility condition), derivatives of a Gaussian function and many others. It is actually often relevant to consider so-called analytic wavelets, i.e. complex wavelets whose Fourier transform vanishes for negative frequencies. Among these, popular choices include the so-called Morse wavelets~\cite{Olhede2002generalized}, defined by their Fourier transform $\hat\psi(\nu)=K \nu^k e^{-\nu^\gamma}$ for $\nu >0$, the time-frequency localization of which is controlled by the parameters $k\in\NN^*$ and $\gamma\in\RR_+^*$ ($K$ is a normalization constant) and the Altes wavelet~\cite{Altes1973some}. We will focus here on the so-called sharp wavelet, used in~\cite{Meynard2018spectral}
\begin{equation}
\hat\psi(\nu) = \exp\left(\ln(\epsilon)\frac{\delta(\nu,\nu_0)}{\delta(\nu_1,\nu_0)}\right)\ ,\quad\nu\in\RR_+\ .
\end{equation}
Here $\nu_0$ is the mode of $\hat\psi$  and $\nu_1$ is a cutoff frequency, defined so that $\hat\psi(\nu_1)=\epsilon$ for some prescribed tolerance $\epsilon<1$. The divergence $\delta$ is defined by $\delta(a, b) = \frac{1}{2}\left(\frac{a}b + \frac{b}a \right)- 1$. The quality factor of $\psi$, i.e. the center frequency to bandwidth ratio, can be expressed
as $Q = 1/\sqrt{C(C + 4)}$, where $C = -\delta(\nu_1 , \nu_0 ) \ln (2)/ \ln (\epsilon) > 0$.

Using such wavelets, the wavelet transform $\cW_x$ of real valued signals is complex valued. The synthesis formula~\eqref{fo:wavelet.synthesis} has to be modified accordingly, to read (for a suitably chosen normalization of $\psi$)
\begin{equation}
\label{fo:wavelet.synthesis.real}
x = \Re\left(\int_{-\infty}^\infty \psi_s*\cW_x(s,\cdot)\,\frac{ds}s\right) \ .
\end{equation}

\medskip
Since we limit to time-scale representation, we follow the standard logarithmic discretization scheme for the scale variable $q^s$, equivalently linear discretization scheme for $s$. Time is sampled on a regular, scale independent lattice. The discrete wavelet transform used throughout this paper is therefore very close to the so-called stationary wavelet transform (except that scales are not necessarily integer powers of 2).

\subsection{JEFAS: an analysis-based approach}
The current paper is mainly devoted to a synthesis-based approach, however it also relies on an earlier approach called JEFAS (Joint Estimation of Frequency, Amplitude and Spectrum) which we outline here for completeness.
\subsubsection{Background}
Let $Y$ denote a signal generated by time warping from some unknown signal $X$, as in~\eqref{fo:time.warping.rand}. $X$ is assumed to be zero mean and wide sense stationary, with power spectral density $\cS_X$ distributed following a normal law

The approach relies on approximations of the following form~\cite{Meynard2018spectral}. Assume polynomial decay for the wavelet: $|\psi(t)|\le (1+|t|^\beta)\inv<\infty$ for some $\beta>2$. Assume further that the power spectrum $\cS_X$ of the underlying stationary signal $X$ is such that \edited{$\int_0^\infty \nu^{2\rho}\cS_X(\nu)\,d\nu<\infty$} with $\rho=(\beta-1)/(\beta+2)$. Then we have
\begin{equation}
\label{fo:JEFAS}
\cW_Y(s,\tau) = \alpha(\tau)\cW_X(s + \edited{\log_q}(\gamma'(\tau)), \gamma(\tau)) + R(s,\tau) = \widetilde{\cW}_Y(s,\tau) + R(s,\tau)\ .
\end{equation}
\edited{Here $\widetilde{\cW}_Y$ is the approximate transform, and $R$ is a remainder. In a nutshell, amplitude modulation and time warping in the time domain amount, up to some accuracy, to amplitude modulation and translation in the time-scale domain.

For completeness, we sketch here the proof of the result, which is given in details in~\cite{Meynard2018spectral} (a deterministic version can be derived similarly following the lines of~\cite{Clerc2003estimating}). The idea is that whenever the functions $\alpha$ and $\gamma$ are smooth enough, they can be Taylor-expanded around a time value of interest: 
given a test function $g$ located near $t=\tau$ (\ie decaying fast enough as a function of $|t-\tau|$), Taylor expansions near $t=\tau$ yield
\begin{align*}
\cM_\alpha g(t)&\approx\cM_{\alpha(\tau)} g(t)\ ,\\
\cD_\gamma g(t)&\approx\widetilde{\cD_\gamma^\tau} g(t)\,,\quad  \hbox{with}\quad \widetilde{\cD_\gamma^\tau} \defeq T_\tau \cD_{\!-\log_q\!(\!\gamma'(\tau)\!)}T_{\!-\gamma(\tau)}\, 
\end{align*}
$T_\tau$ denoting translation by $\tau$. 
Therefore, we approximate the wavelet transform of $Y$ by
\[
\cW_Y(s,\tau)\approx \left\langle \cM_{a(\tau)}\widetilde{\cD_\gamma^\tau} X,T_\tau \cD_s\psi\right\rangle
= \alpha(\tau)\left\langle \cD_{\!-\log_q\!(\!\gamma'(\tau)\!)}T_{-\gamma(\tau)}X,\cD_s\psi\right\rangle
\]
which yields the desired expression since the adjoint operators of $\cD$ and $T$ are such that $\cD_s^*=\cD_{-s}$ and $T_\tau^* = T_{-\tau}$. Theorem 1 in~\cite{Meynard2018spectral} provides upper bounds for the remainder term $R$.}

\medskip

Provided the remainder $R$ in~\eqref{fo:JEFAS} can be neglected, $\cW_Y$ may be approximated by $\widetilde{\cW}_Y$, which is a circular (complex) Gaussian random field (see~\cite{Picinbono1996second} for details on circular normal distributions, the main aspects are sketched in~\ref{app:complex.gaussian}). The distribution of $\widetilde{\cW}_Y$ is characterized by its covariance kernel $C_Y$, which may be written in terms of the covariance $C_X$ of $X$, $\alpha$ and $\gamma$. This suggests to estimate the unknown parameters using maximum likelihood (ML) approach. Unfortunately, ML is intractable in this case, as it leads to huge nonlinear optimization problems. The JEFAS algorithm of~\cite{Meynard2018spectral} proposes an ansatz that reduces significantly the computational load: fixed time slices of the wavelet transform $\cW_Y$ are considered independent.

\subsubsection{Estimation}
\label{ssse:JEFAS.estimation}
Suppose one is given a finite dimensional wavelet transform $\underline{\bW}\in\CC^{M_s\times N_\tau}$, and denote by $\bW=\mathsf{vec}(\underline{\bW})\in \CC^{M_s N_\tau}$ the vectorized version. $\underline{\bW}$ is assumed to result from the sampling of the continuous transform $\cW_Y$ of a time warped signal $Y$ as in~\eqref{fo:time.warping.rand}, on a time-scale grid $\Lambda = M_s\times N_t$.  The goal is to estimate the set of parameters $\Theta = (\theta_1,\theta_2,\theta_3 )$, where $\theta_1,\theta_2,\theta_3\in\RR^{N_\tau}$ are defined by $\theta_1( \tau_n)=\alpha(\tau_n)^2$ , $\theta_2( \tau_n)=\log_q (\gamma'(\tau_n))$ and $\theta_3( \tau_n)=\gamma(\tau_n)$. Here $\tau_n$ denotes the $n$-th time sample, for $n = 1\dots N_\tau$. Parameters corresponding to time $\tau_n$ are denoted by $\Theta_n$.

As alluded to above, consider fixed time slices $\bw_n$, $n=1\dots N_\tau$ of $\bW$ as independent random vectors leads to approximating the covariance matrix $\bC_\bW(\Theta)$ by a block diagonal matrix, each block $\bC_n(\Theta_n)$ corresponding to a fixed time index $n$. The negative log likelihood $\cB(\bTheta) = -\cL(\bTheta)$ takes the form of a Bregman $\log\det$ divergence~\cite{Cherian2013jensen}  between the covariance matrices $\bW\bW^H$ and $\bC(\bTheta)$ (up to an additive constant that depends on $\bW$). As a consequence of the independence assumption, $\cB(\bTheta)$ can be approximated by a sum of reduced (i.e. fixed time) terms $\cB_r(\Theta_n)$, which can be optimized independently
\begin{eqnarray}
\label{fo:JEFAS.Bregman}
\cB(\Theta) &=&  \mathsf{Tr}\left(\bW\bW^H \bC_\bW(\Theta)\inv\right)- \ln\det \left(\bC_\bW(\Theta)^{-1}\right)\\
\label{fo:JEFAS.Bregman2}
&\approx& \sum_{n=1}^{N_\tau} \cB_r(\Theta_n)\ ,\qquad\text{where}\\
\label{fo:JEFAS.Bregman3}
\cB_r(\Theta_n) &=&  \Tr{\bw_n\bw_n^H \bC_n(\Theta_n)^{-1}}  - \ln\det\left(\bC_n(\Theta_n)^{-1}\right)
\end{eqnarray}
Notice that the covariance matrix $\bC_n(\Theta_n)$ also depends on the unknown power spectrum $\cS_X$ of the underlying stationary signal $X$, its matrix elements are given as follows: for each pair of scales $i,j=1,\dots M_s$,
\begin{equation}
\label{fo:JEFAS.cov}
\left(\bC_n(\Theta_n)\right)_{ij} = \theta_{1}(\tau_n)\, f(s_i+\theta_2(\tau_n),s_j+\theta_2(\tau_n))\ ,
\end{equation}
where $f$ is a Hermitian symmetric, positive semi-definite function defined by
\begin{equation}
\label{fo:JEFAS.cov.2}
f(s,s') = q^{(s+s')/2}\int_0^\infty\cS_X\left(\xi\right)
\overline{\hat\psi}\left(q^{s}\xi\right)\hat\psi\left(q^{s'}\xi\right) \, d\xi \ .
\end{equation}
The function $f$ is called \textit{covariance template} of the model.

\edited{Notice also that the covariance matrix $\bC_n(\Theta_n)$ does not depend explicitly on parameters $\theta_3(\tau_n)$ anymore. This is a consequence of the independence assumption made on fixed-time slices of the time-scale transform.}

Therefore, estimating the parameters requires the knowledge of the power spectrum $\cS_X$. Given current values of the parameters, an estimate $\widetilde{\cW_X}$ of the wavelet transform $\cW_X$ of the underlying stationary process can be obtained by interpolation using~\eqref{fo:JEFAS}, from which an estimate of the power spectrum can be computed by time averaging $|\widetilde{\cW_X}|^2$ (see Chapter 6 in~\cite{Carmona1998practical}).

\smallskip
The JEFAS algorithm proceeds by iterative updates of the parameter $\Theta$ and power spectrum $\cS_X$. At convergence, estimates for the time warping $\gamma$ and amplitude modulation $\alpha$ are finally obtained by interpolation from the parameters $\Theta$. Details of the algorithm can be found in Algorithm 1 in~\cite{Meynard2018spectral}, which also provides statistical performance assessments.

\subsubsection{Discussion}
We now discuss some pros and cons of JEFAS.

We believe JEFAS provides a relevant approach to extend the spectral estimation problem to non-stationary situations that are well accounted for the the AM-TW (amplitude modulation combined with time warping) model. While an algorithm similar to JEFAS can also be developed for AM-FM models, time warping is probably a more relevant model than frequency modulation in a number of situations of interest.

JEFAS is rooted on an approximate maximum likelihood approach, and therefore  benefits from tools of the statistical machinery. This includes among others precision estimates (see~\cite{Meynard2018spectral,Meynard2019stationnarites}), which are useful to assess quantitatively the performances of the approach.  

However, JEFAS is mainly a post-processing of a time-frequency transform (here a wavelet transform), and thus inherits from the weaknesses of the latter. These include time-frequency resolution limited by classical time-frequency uncertainty, and a difficulty to handle fast variations of the time-dependent spectrum. In addition, JEFAS is based upon simplifying assumptions, in particular assumptions of slow variations of the time warping function. This increases the difficulty of JEFAS to handle fast non-stationarities.

These weaknesses motivate the development of an alternative approach, that avoids a prior computation of time-frequency transform and explicit slow variations assumptions. This approach is described in the next section.

\section{JEFAS-S: a time warping based signal synthesis model}
\label{se:JEFAS-S}
We now turn to the main contribution of this paper, namely the synthesis-based approach. Instead of starting from a transform, the approach is based on a synthesis model for a non-stationary random signal $Y$, in the form
\begin{equation}
\label{fo:JEFAS-S}
Y = \Re\left(\sum_{s\in S}  \psi_s * W_s \right)+ \varepsilon\ ,
\end{equation}
where $\varepsilon$ is a zero mean, Gaussian white noise with variance $\sigma^2$. For each scale $s$, $W_s$ is a random process, with prescribed prior probability distribution, which will implement the time warping model. Based on the latter, the goal is to estimate $W_s$ from a realization $\by$ of $Y$ for all scales $s$ under consideration.

\subsection{Problem formulation in finite dimensional setting}
As before, we denote by $(\tau_1,\dots \tau_{N_\tau})$ the vector of times at which the signal is sampled. Signals of interest and wavelets $\psi_s$ are then $N_\tau$-dimensional vectors, for which we further assume periodic boundary conditions: $x(\tau_{n+N_\tau})=x(\tau_n)$.

Let $\by\in\RR^{N_\tau}$ be the signal under consideration. We will seek estimates for $W_s$ at time samples.
We denote by $S=\{1,2,\dots M_s\}$ the set of considered scales. For each $n=1,\dots N_\tau$, let $\bw_n\in\RR^{M_s}$ denote the vector of wavelet coefficients at time $\tau_n$, and let $\ubW =\begin{pmatrix}\bw_1&\dots&\bw_{N_\tau}\end{pmatrix}$ be the resulting wavelet transform matrix. For every $n=1,\dots N_\tau$, introduce the matrix
\begin{equation}
\bPsi_n = \begin{pmatrix}
\psi_{s_1}(\tau_{1-n})&\psi_{s_2}(\tau_{1-n})&\dots&\psi_{s_{M_s}}(\tau_{1-n})\\
\psi_{s_1}(\tau_{2-n})&\psi_{s_2}(\tau_{2-n})&\dots&\psi_{s_{M_s}}(\tau_{2-n})\\
\vdots&\vdots&\ddots&\vdots\\
\psi_{s_1}(\tau_{N_\tau-n})&\psi_{s_2}(\tau_{N_\tau-n})&\dots&\psi_{s_{M_s}}(\tau_{N_\tau-n)}
\end{pmatrix}
\in\CC^{N_\tau\times M_s}\ ,
\end{equation}
then equation~\eqref{fo:JEFAS-S} reads
\begin{equation}
\label{fo:JEFAS-S.2}
\by = \Re\left(\sum_{n=1}^{N_\tau} \bPsi_n\bw_n\right)+\beps  = \Re\left(\bD\bW\right)+\beps\ ,
\end{equation}
where we have introduced the matrices $\bD\in\CC^{N_\tau\times N_\tau M_s}$ and $\bW\in\CC^{N_\tau M_s}$ defined by
\begin{equation}
\bD = \begin{pmatrix}\bPsi_1&\bPsi_2&\dots&\bPsi_{N_\tau}\end{pmatrix}\ ,\qquad
\bW = \mathsf{vec}(\ubW)=\begin{pmatrix}\bw_1\\\bw_2\\\vdots\\\bw_{N_\tau}\end{pmatrix}\ .
\end{equation}

\subsection{Random model for noise and time-scale coefficients}
\label{sse:JEFAS-S.model}
As mentioned above, the noise $\varepsilon$ in~\eqref{fo:JEFAS-S} is modelled as a real Gaussian white noise, with zero mean and variance $\sigma^2$, therefore $\beps\sim \cN(0,\sigma^2\,\Id_{N_\tau})$.

We now introduce the time-scale warping prior model on $\bW$. This prior model is designed so as to mimic the behavior of wavelet transform of time warped stationary processes, and involve the following elements
\begin{enumerate}
\item 
Columns $\bw_n$ are assumed independently distributed, according to a circular complex Gaussian law (this was an approximation in JEFAS, and is now a prior model in JEFAS-S). We refer to \textit{e.g.}~\cite{Picinbono1996second} for an extended description of multivariate complex Gaussian distributions, a short account in given in~\ref{app:complex.gaussian}.
\item
The covariance matrix $\bC_n$ of a given column $\bw_n$ takes a form similar to equations~\eqref{fo:JEFAS.cov}, for some Hermitian symmetric, positive semi-definite covariance template $f$. \edited{The choice for $f$ depends on the application and the a priori shape of the underlying stationary signal. Two covariance templates are investigated in this paper.

\begin{enumerate}
\item 
A natural choice for $f$ is given by~\eqref{fo:JEFAS.cov.2}. The only difference is that since amplitude modulation is not considered here, the parameter $\theta_1(\tau_n)$ is not present any more. This prior brings a prescribed regularity of representation along the scale axis. This ensures a good robustness of the representation to errors in estimation of the translation parameter $\theta_2(\tau_n)$.

\item
When the underlying stationary signal is a sum of $K$ sine waves, a priori constraints on the sparse structure of the power spectrum can be made. To do this, we simply modify the expression of the covariance template $f$ by replacing $\cS_X$ in~\eqref{fo:JEFAS.cov.2} by
\begin{equation}
\label{fo:sharp.spectrum}
\cS_{X,\mathrm{sp}}(\xi) \defeq \sum_{k=1}^K a_k^2 \delta(\xi- \xi_k)\,,
\end{equation}
where $a_k$ and $\xi_k$ are the amplitudes and frequencies of the sine waves forming the stationary signal. This choice provides sharper time-scale representations.
\end{enumerate}

}

\end{enumerate}

\edited{
\begin{remark}[Comparison with JEFAS]
\label{rem:JEFASvsJEFASS}
\hfill
\begin{enumerate}
\item
Unlike JEFAS, amplitude modulation is not included in JEFAS-S. This is motivated by the fact that in JEFAS, amplitude modulation estimation was shown to be poorer than time warping. Nevertheless, there is no theoretical obstruction to the inclusion of amplitude modulation in the a priori model if necessary. The estimation of amplitude modulation is briefly addressed in Section~\ref{subsubsec.JEFASS.AM} below.
\item
The property of independence of the columns of the time-scale transform is a simplification in JEFAS, which is actually quite questionable, in particular for large scales. In JEFAS-S, this simplification is implemented as a property of the prior probability distributions, which is less constraining: choosing independent prior means that we don't want to assume that columns are a priori informative about each other, and leave it to the observations to tune these dependences. As we shall see, the posterior distribution of columns does not have this independence property any more.
\end{enumerate}
\end{remark}
}
\medskip
Summarizing, the model parameters are the noise variance $\sigma^2$, the warping parameters $\theta_2(\tau_n)$ and the covariance kernel $f$.

\subsection{Bayesian estimation}
\subsubsection{Posterior expectation}
Let us first assume that the model parameters are known, we want to estimate the vector $\bW$, using the posterior expectation. Since we are in the Gaussian case, the posterior expectation coincides with the maximum a posteriori estimator. 

The prior distribution of $\bW$ is the circular complex Gaussian
\begin{equation}
\label{fo:W.prior}
p(\bW) = \frac{1}{\pi^{N_\tau M_s}\det(\bGamma_\mathsf{pr})} \exp\left(-\frac{1}{2} \bW^H\bGamma_\mathsf{pr}^{-1}\bW\right)
\end{equation}
(see~\ref{app:complex.gaussian} for details) where the prior covariance matrix $\bGamma_\mathsf{pr}$ is the block diagonal matrix with blocks $\bC_n$:
\begin{equation}
\bGamma_\mathsf{pr} = \mathsf{bdiag}\left(\bC_1,\dots \bC_{N_\tau}\right)\in\CC^{N_\tau M_s\times N_\tau M_s}
\end{equation}
As mentioned above, the noise $\beps\sim \cN(0,\sigma^2\,\Id_{N_\tau})$ is white Gaussian and real valued. Therefore the likelihood of observations is given by 
\begin{equation}
\label{fo:obs.likelihood}
p(\bY|\bW) \sim \cN(\Re(\bD\bW),\sigma^2\Id_{N_\tau})
\end{equation}

When both noise and prior are real Gaussian, it is known that the posterior distribution is also real Gaussian. We provide here a similar result with circular complex Gaussian prior, which shows that the posterior distribution is complex Gaussian, non circular.
\begin{proposition}
\label{prop:posterior}
Let $\bW\in\CC^{N_\tau\times M_s}$ be a circular complex Gaussian random vector, distributed following~\eqref{fo:W.prior}. Let $\by\in\RR^{N_\tau}$ be a random vector whose likelihood  $p (\by |\bW )$ is as in~\eqref{fo:obs.likelihood}. Then the posterior distribution $p (\bW | \by )$ is a complex Gaussian law
\begin{equation}
\label{fo:posterior}
p(\bW|\by)\sim\cCN(\bmu_\mathsf{po},\bGamma_\mathsf{po},\bR_\mathsf{po})
\end{equation}
with
\begin{eqnarray}
\label{fo:mu.po}
\bmu_\mathsf{po}&=&\bGamma_\mathsf{pr}\bD^H\bC_\by^{-1} \by \\
\label{fo:Gamma.po}
\bGamma_\mathsf{po}&=&\bGamma_\mathsf{pr}-\frac{1}{4}\bGamma_\mathsf{pr}\bD^H\bC_\by^{-1}\bD\bGamma_\mathsf{pr} \\
\bR_\mathsf{po}&=& -\left(\bD^H\bD + 4\sigma^2 \bGamma_\mathsf{pr}^{-1}\right)^{-1} \overline{\bD^T\bD}\,\overline{\Gamma_\mathsf{po}}
\end{eqnarray}
and
\begin{equation}
\bC_\by =\sigma^2\Id + \frac{1}{2}\Re\left(\bD\bGamma_\mathsf{pr} \bD^H\right)
\end{equation}
\end{proposition}
The proof is given in~\ref{app:prop1.proof}. The expression of the relation matrix $\bR_\mathsf{po}$ is provided for the sake of completeness, although it will not be used in the sequel.\hfill$\square$

\medskip
Let $\tbW = \EE_{\bW\vert\by}\left\{\bW\right\}$ denote the conditional expectation of $\bW$. We have that $\tbW=\bmu_\mathsf{po}$, the posterior mean. The columns $\tbw_n$ of the time-frequency representation $\tbW$ can be written as
\begin{equation}
\label{fo:JEFAS-S.WTestimation}
\tbw_n = \frac{1}{2} \bC_n\bPsi_n^H\bC_\by^{-1}\by\ ,
\end{equation}
where the block diagonal structure of $\bGamma_\mathsf{pr}$ yields the following form for $\bC_\by$
\begin{equation}
\bC_\by = \sigma^2\,\Id_{N_\tau} + \frac{1}{2}\Re\left(\sum_{n=1}^{N_\tau} \bPsi_n\bC_n\bPsi_n^H
\right)\ .
\end{equation}
Notice that $\bC_\by$ is a matrix of dimension $N_\tau\times N_\tau$, which can be large for real world signals, yielding heavy computational load for evaluation, storage and inversion.
\begin{remark}
\edited{As stressed in Remark~\ref{rem:JEFASvsJEFASS}, vectors $\bw_n$ are modeled as independent vectors in the prior model.
However, estimated vectors $\tbw_n$, which take into account the actual signal,} are not uncorrelated anymore:
\[
\Ex{\tbw_n\tbw_{n'}^H} =\delta_{nn'}\bC_n - \frac{1}{4}\,\bC_n\bPsi_n^H\bC_\by^{-1}\bPsi_{n'}\bC_{n'} \ ,
\]
which has no reason to vanish for $n\ne n'$.
\end{remark}

\subsubsection{Parameter estimation: JEFAS-S}
We now turn to the parameter estimation problem. In this section we assume that the covariance template $f$ and the noise variance $\sigma^2$ are fixed.  The choice of $f$ will be discussed later on, and the noise variance $\sigma^2$ will play the role of regularization parameter.

Let us recall the parameters of the AM-TW model, that were introduced in Section~\ref{ssse:JEFAS.estimation}: $\Theta = (\theta_1,\theta_2,\theta_3 )$, with $\theta_1( \tau_n)=\alpha(\tau_n)^2$ , $\theta_2( \tau_n)=\log_q (\gamma'(\tau_n))$ and $\theta_3( \tau_n)=\gamma(\tau_n)$. JEFAS-S only considers the TW model ($\alpha=1$) and, as in JEFAS, the time decorrelation in the prior model implies that only $\theta_2$ enters into the estimation procedure.

For the sake of simplicity, and to avoid confusions with the JEFAS parameters, let us denote here by $\vartheta_n = \theta_2(\tau_n)$, and $\bvtheta=(\vartheta_1,\dots\vartheta_{N_\tau})\in\RR^{N_\tau}$ the remaining parameters to be estimated. With these notations, the covariance matrix associated to time $\tau_n$ takes the form
\begin{equation}
\label{fo:JEFAS-S.cov}
\bC_n = \bC(\vartheta_n)\ ,\quad\text{with}\quad
\left(\bC(\vartheta)\right)_{ij} = f(s_i+\vartheta,s_j+\vartheta)\ ,\ \forall\vartheta\ .
\end{equation}
Here $f$ is a covariance template, as introduced in~\eqref{fo:JEFAS.cov.2}, which will control the behavior of the time-frequency representation $\bW$.

\smallskip
$\bvtheta$ can be estimated using an Expectation-Maximization (EM) strategy~\cite{Dempster1977maximum}.  \edited{EM is an iterative algorithm that can be used to search for the maximum likelihood estimators of model parameters. In general, EM is not guaranteed to converge to a global maximum of the likelihood. However, the likelihood is guaranteed to increase at each iteration.}

After some initialization, parameters $\bvtheta$ are updated iteratively. Denoting by $\tbvtheta^{(k-1)}$ the estimate from iteration $k-1$, a new estimate $\tbvtheta^{(k)}$ is computed as follows
\begin{itemize}
\item 
\textit{Expectation step:} introduce the auxiliary function
\begin{equation}
\label{fo:EM.Q}
Q\left(\bvtheta,\tbvtheta^{(k-1)}\right) = \EE_{(k-1)}\!\left\{\ln\left(p_{\bvtheta}(\by,\bW)\right)\right\}
\end{equation}
where $p_{\bvtheta}(\by,\bW)$ is the joint probability density function of $\bW$ and $\by$, which depends on $\bvtheta$, and where we have denoted by $\EE_{(k-1)}$ the expectation with respect to $\bW$ conditional to $\by$, using parameter value $\tbvtheta^{(k-1)}$. 
\item
\textit{Maximization step:} solve
\begin{equation}
\label{fo:theta.update}
\tbvtheta^{(k)} = \mathop{\mathsf{arg\,max}}_{\bvtheta}\, Q\!\left(\bvtheta,\tbvtheta^{(k-1)}\right)\ .
\end{equation}
\end{itemize}
The application of the EM paradigm leads to the following update rules~\cite{Meynard2019stationnarites}. For the sake of simplicity let us introduce some notations. After iteration $k-1$, denote by $\tbW^{(k-1)}$ and $\tbw_n^{(k-1)}$ the current estimates for the time-scale representation and its columns, and by $\tbC_n^{(k-1)}$ the current estimate of the $n$-th block of $\bC_\by$. Let also $\tbC_\by^{(k-1)}$ denote the corresponding estimate of $\bC_\by$ (thus depending on $\tbvtheta^{(k-1)})$.
Similarly, we denote by $\tbGamma_n^{(k-1)}$ the $n$-th diagonal block of the estimate of the posterior covariance matrix $\bGamma_\mathsf{po}$ from iteration $k-1$ (which is a function of $\tbvtheta^{(k-1)}$). 
\begin{proposition}[EM update rules]
\label{prop:EM.update}
With the above notations, the update rules for the time scale representation and parameters read
\begin{enumerate}
\item
Time scale representation
\begin{equation}
\label{fo:wn.update}
\tbw_n^{(k)} = \frac{1}{2} \tbC_n^{(k-1)}\bPsi_n^H\left(\tbC_\by^{(k-1)}\right)^{-1}\by\ ,
\end{equation}
\item
Parameters:
\begin{eqnarray}
\nonumber
\tbvtheta^{(k)} &=& \mathop{\mathsf{arg\,min}}_{\bvtheta}
\sum_{n=1}^{N_\tau}\bigg[ \ln\det(\bC(\vartheta_n)) + \tbw_n^{(k-1)\,H}\bC(\vartheta_n)^{-1}\tbw_n^{(k-1)}\\
&&\hphantom{aaaaaaaaaaa}+\Tr{\bC(\vartheta_n)^{-1}\Re\left(\tbGamma_n^{(k-1)}\right)}\bigg]
\label{fo:JEFAS-S.optim}
\end{eqnarray}
\end{enumerate}
\end{proposition}
The proof is given in~\ref{app:proof.EM.update}.

\begin{remark}
The parameter update can also be written as
\begin{eqnarray}
\tbvtheta^{(k)}\!\! &\!=\!&\!\! \mathop{\mathsf{arg\, min}}_{\bvtheta\in\RR^{N_\tau}}\ 
\sum_{n=1}^{N_\tau}\cB_n^{(k-1)}(\vartheta_n)\ ,\qquad\text{where for all}\  \vartheta\\
\cB_n^{(k-1)}(\vartheta)\!\! &\!=\!&\!\! \Tr{\!\left(\tbw_n^{(k)}\tbw_n^{(k)H}\!+\tbGamma_n^{(k-1)}\right)\bC(\vartheta)^{-1}\!} - \ln\det\! \left|\bC(\vartheta)^{-1}\!\right| .
\end{eqnarray}
Notice the similarity with equations~\eqref{fo:JEFAS.Bregman}--\eqref{fo:JEFAS.Bregman3}, involving now Bregman type log det divergences between covariance matrices $\bC(\vartheta)$ and $\tbw_n^{(k)}\tbw_n^{(k)H}\!+\tbGamma_n^{(k-1)}$.
\end{remark}

\subsection{Signal synthesis}
After convergence of the estimation algorithm, denote by $\tbw_n$, $\tbC_n$ and $\tbC_\by$ the corresponding estimated quantities.
According to the JEFAS-S model~\eqref{fo:JEFAS-S.2}, the denoised signal given by
\begin{equation}
\by_0 = \Re\left(\sum_{n=1}^{N_\tau} \bPsi_n\bw_n\right)
\label{fo:synthesis}
\end{equation}
can be estimated as
\begin{equation}
\widetilde{\by}_0 = \Re\left(\sum_{n=1}^{N_\tau} \bPsi_n\tbw_n\right)
= \frac{1}{2} \Re\left(\sum_{n=1}^{N_\tau} \bPsi_n\tbC_n\bPsi_n^H\right)\left(\tbC_\by\right)^{-1}\by
\ .
\end{equation}

The quality of the estimation can be assessed by the bias and covariance of the estimator, which are given in the following result.
\begin{proposition}
\label{prop:estimation.assesssment}
Let $\widetilde{\by}_0$ denote the above estimator of the denoised signal.
\begin{enumerate}
\item 
The estimator $\widetilde{\by}_0$ is biased, with bias
\begin{equation}
\bb = \Ex{\widetilde{\by}_0-\by_0} = -\sigma^2\bC_\by^{-1}\by_0\ .
\label{fo:bias}
\end{equation}
\item
The covariance matrix of the estimation error $\be=\widetilde{\by}_0-\by_0$ is given by
\begin{equation}
\bR_\be = \sigma^2 \left(\Id-\sigma^2\bC_\by^{-1}\right)^2\ .
\label{fo:covariance.error}
\end{equation}
\end{enumerate}
\end{proposition}

\begin{remark}
Let us denote by $\gamma_1^2,\gamma_2^2,\dots \gamma_{N_\tau}^2$ the eigenvalues of the Hermitian, positive semidefinite matrix $\frac{1}{2}\Re(\bD\bGammapr\bD^H)$, sorted in increasing order.
\begin{enumerate}
\item
We notice that the bias may be bounded as
\[
\|\bb\|\le \left(1+\frac{\gamma_1^2}{\sigma^2}\right)^{-1}\,\|\by_0\|\ ,
\]
($\gamma_0^2$ is the smallest eigenvalue). The ratio $\gamma_1^2/\sigma^2$ may be interpreted as a signal to noise ratio related to the prior model. The larger the signal to noise ratio, the smaller the bias.
\item
The variance $\Tr{\bR_\be}$ of the estimation error may also be upper bounded, using the submultiplicativity of the trace (i.e. $\Tr{AB}\le\Tr{A}\Tr{B}$), so that
\[
\Tr{\bR_\be}\le N_\tau^2\sigma^2 \left(1-\frac{1}{N_\tau}\sum_{n=1}^{N_\tau}\left(1+\frac{\gamma_n^2}{\sigma^2}\right)^{-1}\right)^2\ .
\]
\end{enumerate}
\end{remark}
\subsection{Algorithm}
\label{sse:algo}
The algorithm whose pseudo-code is given in Algorithm~\ref{alg:JEFAS.adap} implements the approach described above. The algorithm is iterative and therefore requires an initialization and a stopping criterion, which are described below. The input includes the input signal $\by$, the noise variance $\sigma^2$ and a parameter $\varrho$ controlling the stopping criterion. The algorithm also involves a dimension reduction step, which speeds up some matrix operations. This step depends on a parameter  $N'_\tau\ll N_\tau$, which is discussed below.

\begin{algorithm}[t]
\caption{$(\tbW,\tbvtheta,\widetilde{\cS_X})=\operatorname{JEFAS-S}(\by,\sigma^2,\varrho,N'_\tau,f)$}
\begin{algorithmic}[1]
\Require Initial estimates of $\tilde\bvtheta^{(0)}$ and $\tilde\cS_X^{(0)}$ using JEFAS.
\Statex
\State $k\leftarrow 1$.
\While{stopping criterion~\eqref{fo:jefass.stopcrit} = \textsf{FALSE}}
\ForAll{$n\in\{1,\ldots,N_\tau\}$}
\State Restrict $\bPsi_n$, $\bC_y\left(\tbvtheta^{(k-1)}\right)$ and $\by$ to the interval  $[n-N'_\tau/2,n+N'_\tau/2]$.
\State Evaluate $\tbw_n^{(k)}$ using the corresponding reduced version of~\eqref{fo:wn.update}.
\EndFor
\State Estimate $\tbvtheta^{(k)}$ by solving~\eqref{fo:theta.update}.
\State Estimate $\widetilde{\cS}_X^{(k)}$ using the wavelet-based estimate.
\State  Update matrices $\bC_n$ and $\bGammapr$ using $\tbvtheta^{(k)}$, $\widetilde{\cS}_X^{(k)}$ and $f$.
\State $k\leftarrow k+1$.
\EndWhile
\end{algorithmic}
\label{alg:JEFAS.adap}
\end{algorithm}

\subsubsection{Initialization}
The algorithm requires an initial guess $\widetilde{\cS}_X^{(0)}$ for the power spectrum of the underlying stationary signal and for the parameters $\widetilde{\bvtheta}^{(0)}$. These can be obtained by running JEFAS on the input signal $\by$.

\edited{
\begin{remark}
In general, EM is not guaranteed to converge to a global maximum of the likelihood.
However, the likelihood is guaranteed to increase at each iteration. When multiple local minima are expected, the initialization plays a crucial role. 

In our experiments, the initialization described above turned out to provide satisfactory results. When JEFAS fails to converge, an alternative initialization can be obtained by computing at each time a \textit{local scale} defined as a weighted average of the scale, the weights being given by the time-scale transform.

\end{remark}
}
\subsubsection{Stopping criterion}
EM ensures that the likelihood monotonically increases during iterations. The algorithm stops at iteration $k$ when the condition
\begin{equation}
\label{fo:jefass.stopcrit}
\mathcal{L}(\bvtheta^{(k)})-\mathcal{L}(\bvtheta^{(k-1)})<\varrho
\end{equation}
is true. Here $\varrho>0$ is a parameter fixed by the user.

\subsubsection{Dimension reduction}
The algorithm requires solving at each iteration a high dimensional linear system in~\eqref{fo:wn.update}. However, the large matrix $\bC_\by$ turns out to be strongly concentrated around its diagonal. This can be exploited to speed up the evaluation of $\tbw_n$ in~\eqref{fo:JEFAS-S.WTestimation} by restricting $\by$ on a neighborhood $[n-N'_\tau/2,n+N'_\tau/2]$ of the time index $n$ with a certain bandwidth $N'_\tau\ll N_\tau$. The latter has to be matched to the effective support size of wavelets at the largest scale upon consideration.

\subsubsection{Power spectrum estimation}
JEFAS-S requires the knowledge of the power spectrum $\cS_X$ of the underlying stationary signal, both for initialization and during the iterations.

Concerning the initialization, an estimate can be obtained using the JEFAS approach, if the latter is successful. Otherwise, a very crude estimate can be computed using a Welch periodogram from the input signal $\by$.

During the iterations of Algorithm~\ref{alg:JEFAS.adap}, the power spectrum is re-estimated by

using the two-step procedure
\begin{enumerate}
\item
Correcting the time-scale transform $\bW^{(k-1)}$ for scale translations $\vartheta_n^{(k-1)}$
\item
Estimating the wavelet spectrum~\cite{Carmona1998practical}, i.e. the time-average of the corrected time-scale representation.
\end{enumerate}

\edited{
\subsubsection{Amplitude modulation estimation}
\label{subsubsec.JEFASS.AM}
If amplitude modulation is taken into account, the a priori covariance template takes the form~\eqref{fo:JEFAS.cov}. The whole procedure described in the above remains true. It is only necessary to add a step for the update of the estimate of the amplitude modulation parameter between lines 7 and 8 of Algorithm~\ref{alg:JEFAS.adap}. 

Solving problem~\eqref{fo:JEFAS-S.optim} with respect to the amplitude modulation parameter yields a closed-form expression (see eq. (17) in~\cite{Meynard2018spectral}). Therefore, considering the full AM-TW model instead of the TW model alone is computationally inexpensive. As mentioned in Remark~\ref{rem:JEFASvsJEFASS}, amplitude modulation is not considered here due to the poor performance of the AM estimator, which can be assessed in terms of the variance of the estimator. In~\cite{Meynard2018spectral} a closed-form expression was given for the corresponding Cram\'er-Rao lower bound, which was shown to be quite large in general, impairing the quality of estimates.
}

\section{Numerical results}
\label{se:NumRes}

The numerical results presented in the current section are reproducible. The MATLAB code is available online at \url{https://github.com/AdMeynard/JEFAS}.

\subsection{Illustration on wideband non-stationary signal}

\begin{figure}
\centering
\includegraphics[width=\textwidth]{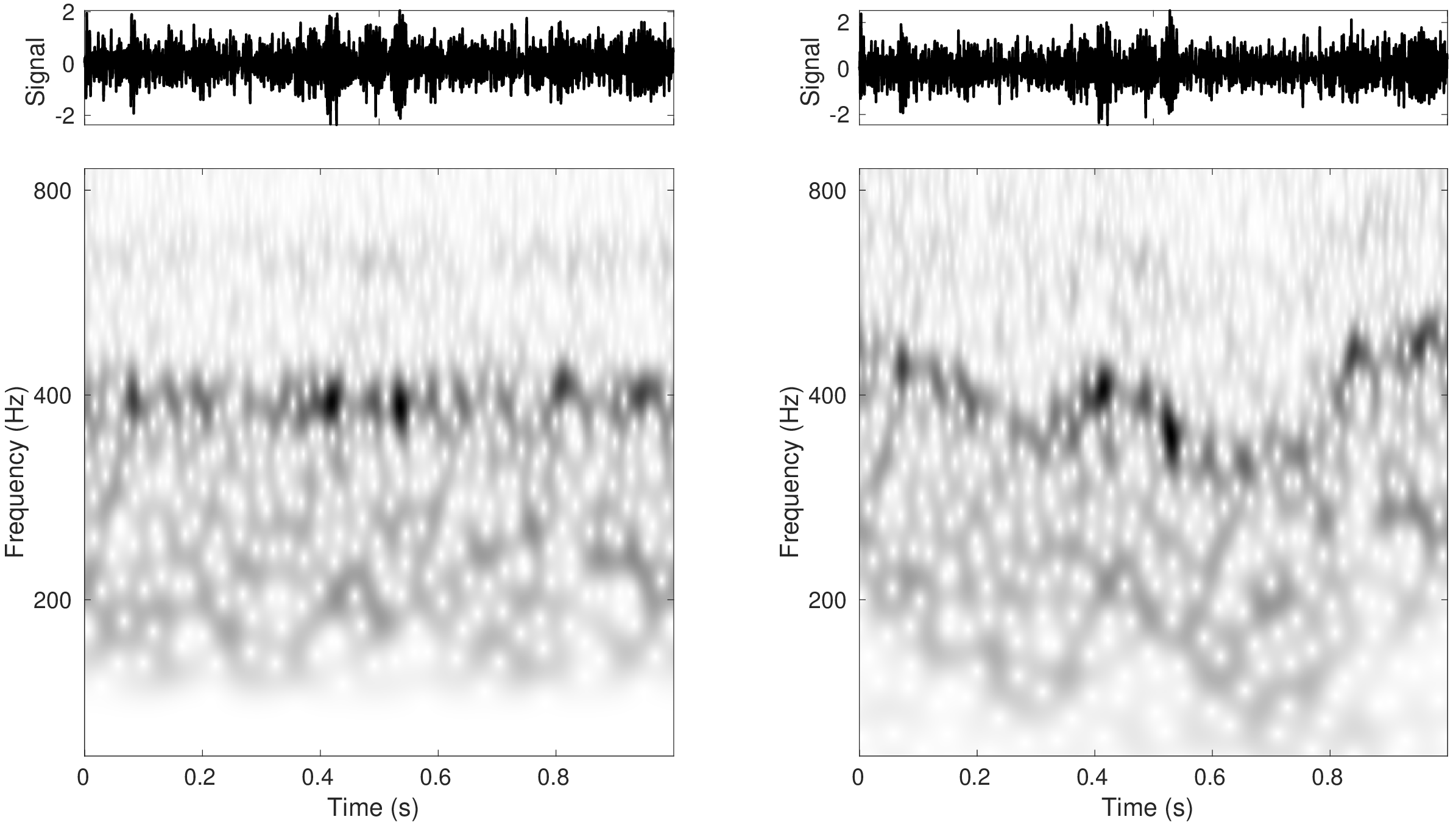}
\caption{Synthetic stationary signal (upper left) and corresponding TW signal (upper right), derived from an audio recording of the wind. Their scalograms are below.}
\label{fi:wind.jefass}
\end{figure}

We implement JEFAS-S on an AM-TW synthesized signal. The underlying stationary signal is formed by a stationary Gaussian random vector of length $N_\tau=2048$. The power spectrum was chosen as the output of JEFAS applied to the audio wind recording whose scalogram is displayed in Fig.~\ref{fi:wind}. The time warping function was also derived from the output of JEFAS on the real signal. Amplitude modulation was not taken into consideration. Eventually, a Gaussian white noise of variance $\sigma^2=0.04$ was added to the observations. The stationary and TW signals, and their respective wavelet transforms are shown in Fig.~\ref{fi:wind.jefass}.

\begin{figure}
\centering
\includegraphics[width=\textwidth]{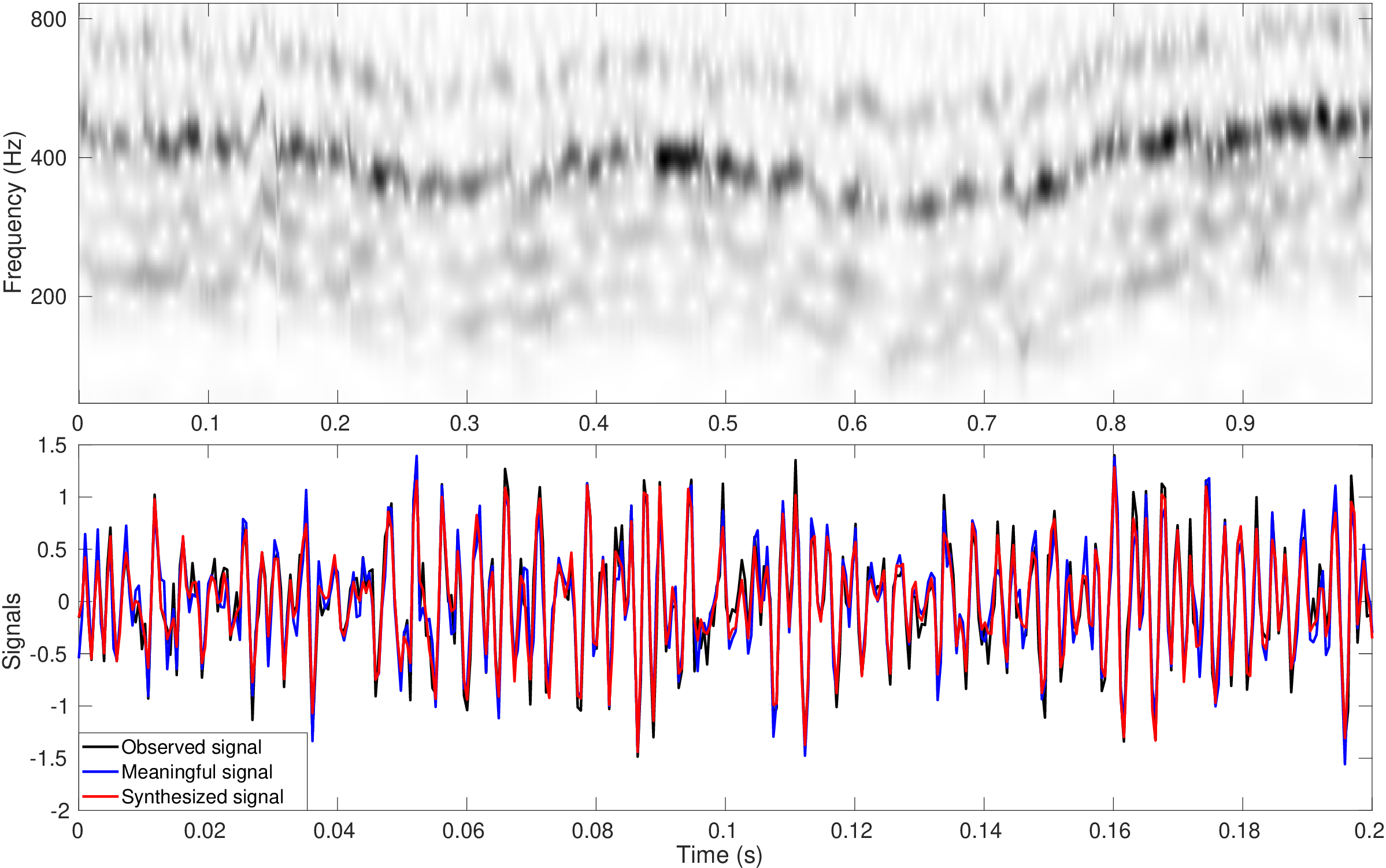}
\caption{JEFAS-S time-scale representation (top), and a segment of the derived synthesized signal (bottom, red) compared to the observations (black) and the noise-free signal (blue).}
\label{fi:synthesis}
\end{figure}

As reported in~\cite{Meynard2018spectral}, slow varying AM-TW signals are satisfactorily processed using JEFAS. Indeed, time warping is well described in the wavelet domain as a time-dependent scale shift of the coefficients. It is therefore estimable via JEFAS. The time warping function is accurately estimated after $6$ iterations: the mean square error is $0.00155$. The initialization of JEFAS-S with the JEFAS output results in a fast convergence in $6$ iterations. The mean square error is $0.00150$. It is incrementally improved at the cost of a significant increase in computing time: $33$~s with JEFAS against $157$~s with JEFAS-S (experiments performed on a laptop equipped with an Intel Core i9 CPU). However, in addition to the estimates of the parameters characterizing the AM-TW model, JEFAS-S provides a time scale representation and a synthesis operator for the noise-free signal. Both quantities are displayed in Fig.~\ref{fi:synthesis}. Visual inspection shows that the time-scale representation provided by JEFAS-S displays more clearly the spectral content coming from the AM-TW signal than the wavelet transform. For example, the frequency bands located around $200$~Hz and $600$~Hz appear clearly, which was not the case in Fig~\ref{fi:wind.jefass}. Moreover, the synthesis operator makes it possible to recover the meaningful part of the signal by filtering out the background noise.

\begin{figure}
\centering
\includegraphics[width=\textwidth]{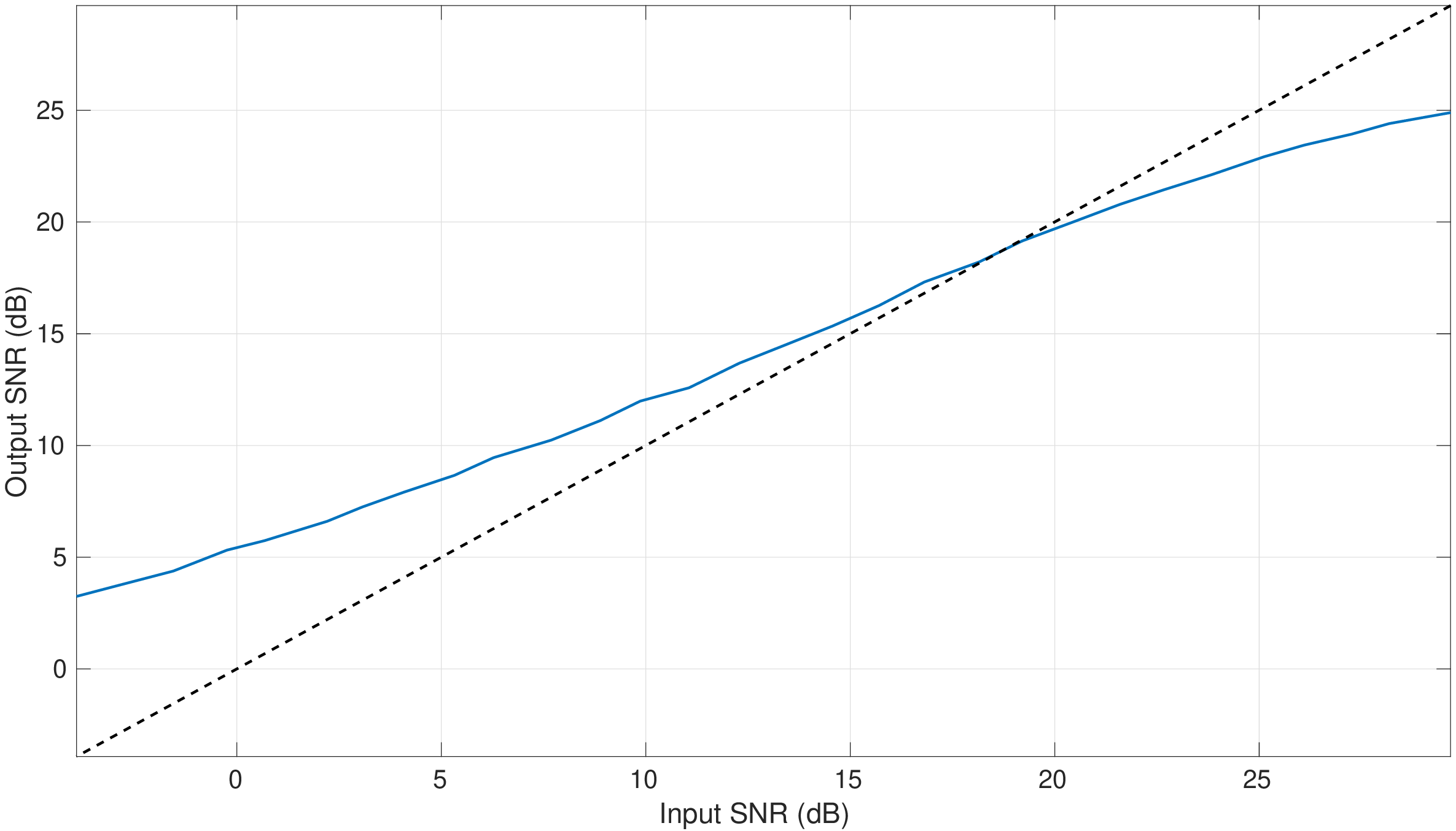}
\caption{Output SNR given by the synthesis estimator, plotted against the input SNR. The dotted line marks the separation between denoising (upper part) and signal degradation (lower part).}
\label{fi:denoising}
\end{figure}

To precisely quantify the ability of JEFAS-S to achieve denoising on AM-TW signals, we apply the synthesis operator to this AM-WP signal with various noise levels ($\sigma^2$ ranging from $0.0005$ to $1$). The input signal-to-noise ratio (SNR) is defined as 
\begin{equation}
\mathrm{SNR_{I}} = 10\log_{10}\left(\dfrac{\int\cS_X(\xi)d\xi}{\sigma^2}\right) .
\end{equation}
The reconstruction formula from the time-scale representation, expressed in~\eqref{fo:synthesis}, provides a supposedly denoised signal. Hence, we measure the output SNR given by
\begin{equation}
\mathrm{SNR_{O}} =  10\log_{10}\left(\dfrac{\EE\{\|\by_0\|^2\}}{\EE\left\{\|\tilde\by_0-\by_0\|^2\right\}}\right)\ .
\end{equation}
In this example, the quality of denoising only depends on the performance of the synthesis estimator determined by its bias~\eqref{fo:bias} and its variance~\eqref{fo:covariance.error}. Indeed, we have $\EE\left\{\|\tilde\by_0-\tilde\by_0\|^2\right\} = \Tr{\bR_\be}+\bb^T\bb$. The output SNR is plotted against the input SNR in Fig.~\ref{fi:denoising}. Numerical results show that for input SNRs smaller than $19$~dB, JEFAS properly performs denoising. When the input SNR is higher than $19$~dB, the intrinsic limitations of the Bayesian estimator become preponderant. The synthesized signal is thus degraded. However, the smaller the input SNR, the better the denoising. The a priori knowledge of the non-stationarity class to which the signal belongs makes the denoising operation adaptive, which can be seen as a form of non-stationary filtering.

The subsequent illustration shows that JEFAS-S is also able to handle AM-TW signals on which JEFAS fails.

\subsection{Illustration on a narrowband non-stationary signal}

We illustrate here the ability of JEFAS-S to handle non-stationary TW signals with fast varying spectral content.

\begin{figure}
\centering
\includegraphics[width=\textwidth]{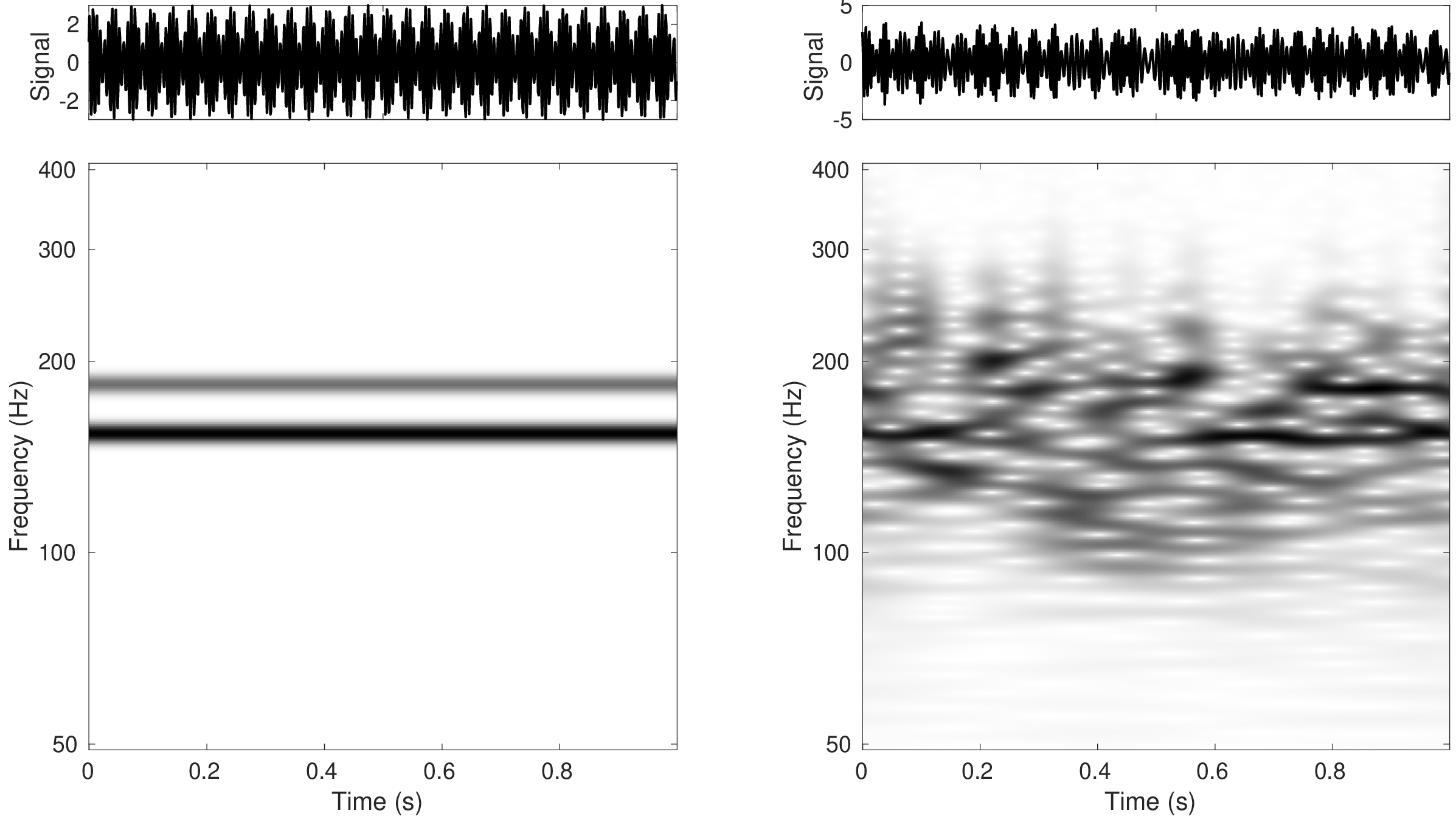}
\caption{Synthetic stationary signal (upper left) and corresponding TW signal (upper right), derived from heartbeat measurements. Their scalograms are below.}
\label{fi:semi.real}
\end{figure}

To this end, a non-stationary signal is synthesized from a stationary signal $\bx$ comprised of two sine waves with close frequencies. We have
\begin{equation}
\bx[n] = \cos\left(2\pi f_1 \frac{n}{N_\tau} + \varphi_1\right) + 2\cos\left(2\pi f_2 \frac{n}{N_\tau}+\varphi_2\right)\ ,
\end{equation}  
where $N_\tau=1024$, $f_1=184$ and $f_2=154$. The phases $\varphi_1$ and $\varphi_2$ are independent random scalars drawn from a uniform distribution in the interval $[0,2\pi)$.  We then apply the time warping operator. The warping function $\gamma$ is derived from the electrocardiogram of a patient with irregular heartbeat. Its derivative is obtained evaluating the R-peak to R-peak intervals, followed by a cubic spline interpolation. This fast varying function is shown in Fig.~\ref{fi:fast.tw.function}. Eventually, a Gaussian white noise of variance $\sigma^2=2.5\times 10^{-3}$ is added to the observations. The stationary and TW signals, and their respective wavelet transforms are displayed in Fig.~\ref{fi:semi.real}.

\subsubsection{JEFAS and JEFAS-S}
\begin{figure}
\centering
\includegraphics[width=\textwidth]{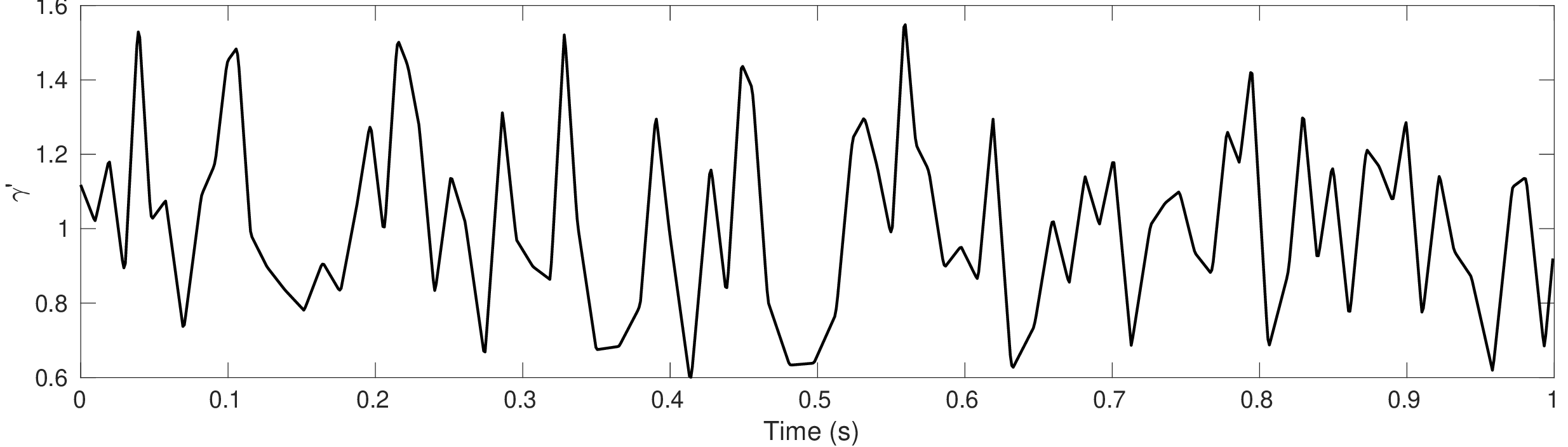}
\includegraphics[width=\textwidth]{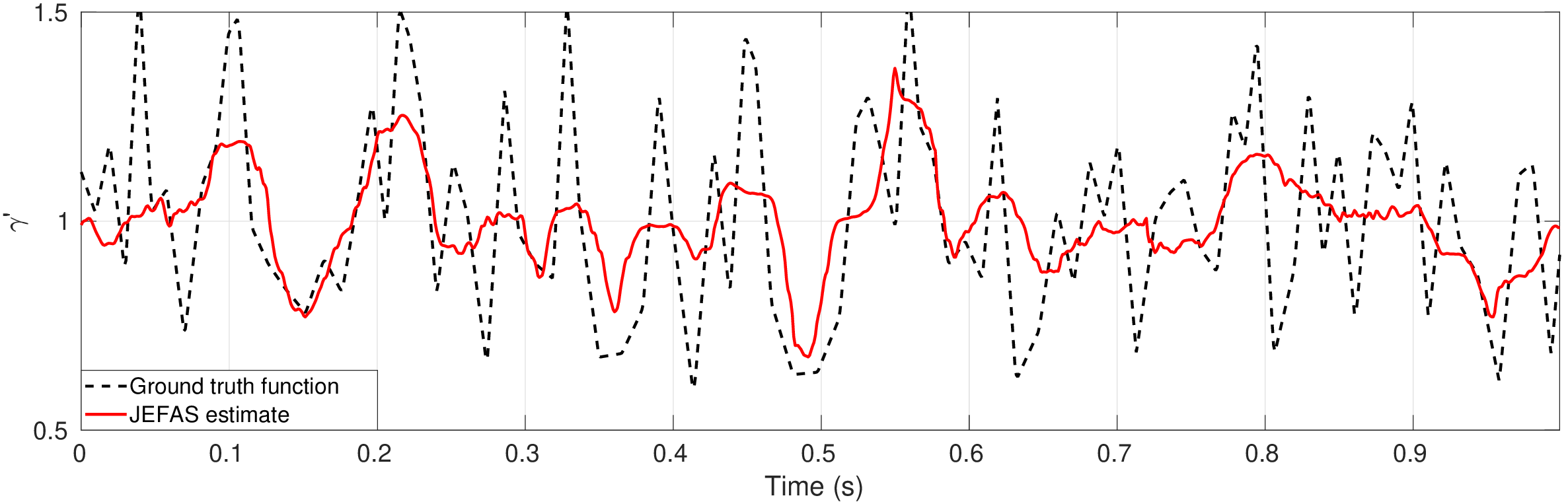}
\includegraphics[width=\textwidth]{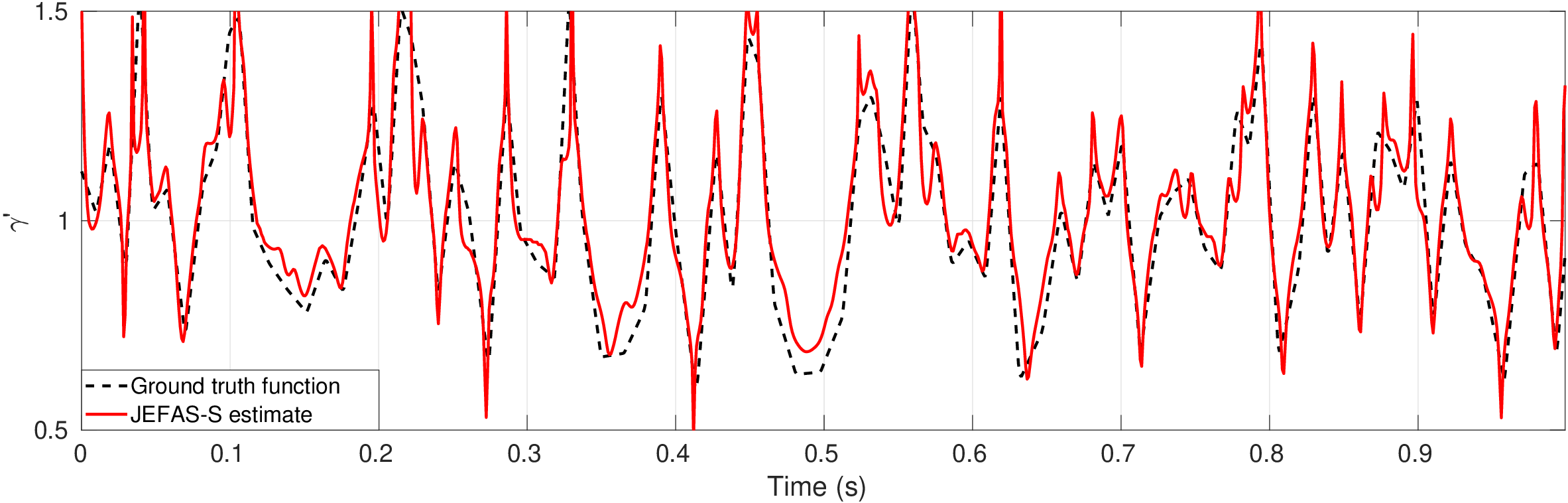}
\caption{Derivative of the warping function, derived from real heartbeat measurements (top). The resulting JEFAS (middle) and JEFAS-S (bottom) estimates are superimposed.}
\label{fi:fast.tw.function}
\end{figure}

The wavelet transform of the TW signal is highly disturbed by interference between the two components. Indeed, on the one hand, the ability to discriminate two close sine waves requires the analysis wavelet to be sharply localized in frequency, but then interference appears along the time axis. On the other hand, the fast variations of the warping function are only discernible if the wavelet is sharp in time, but interference then appears along the scale axis. The impossibility to avoid interference is a consequence of the uncertainty principle. Hence, the approximated behavior of the wavelet transform JEFAS is based on is not valid. On this example JEFAS turns out to converge in $3$ iterations, but the algorithm does not converge satisfactorily towards the ground truth warping function. This is shown in the middle of Fig.~\ref{fi:fast.tw.function}. The rapid variations of $\gamma'$ are strongly smoothed by the JEFAS estimate.

By setting $\varrho=0.5$, JEFAS-S converges in $111$ iterations. The estimated $\gamma'$ function is represented in the bottom of Fig.~\ref{fi:fast.tw.function}. The rapid variations are now accurately reproduced. The accuracy of the estimate is markedly improved: the mean square error goes from $0.0261$ for JEFAS to $0.0069$ for JEFAS-S. The time-scale representation estimated with JEFAS-S is depicted in the top of Fig.~\ref{fi:comp.sst}. Although the presence of two components is not visible, we clearly visualize the rapid variations of the time warping function. Unlike the wavelet transform, interference is absent here.

\subsubsection{Sharpened time-scale representation}
\begin{figure}
\centering
\includegraphics[width=\textwidth]{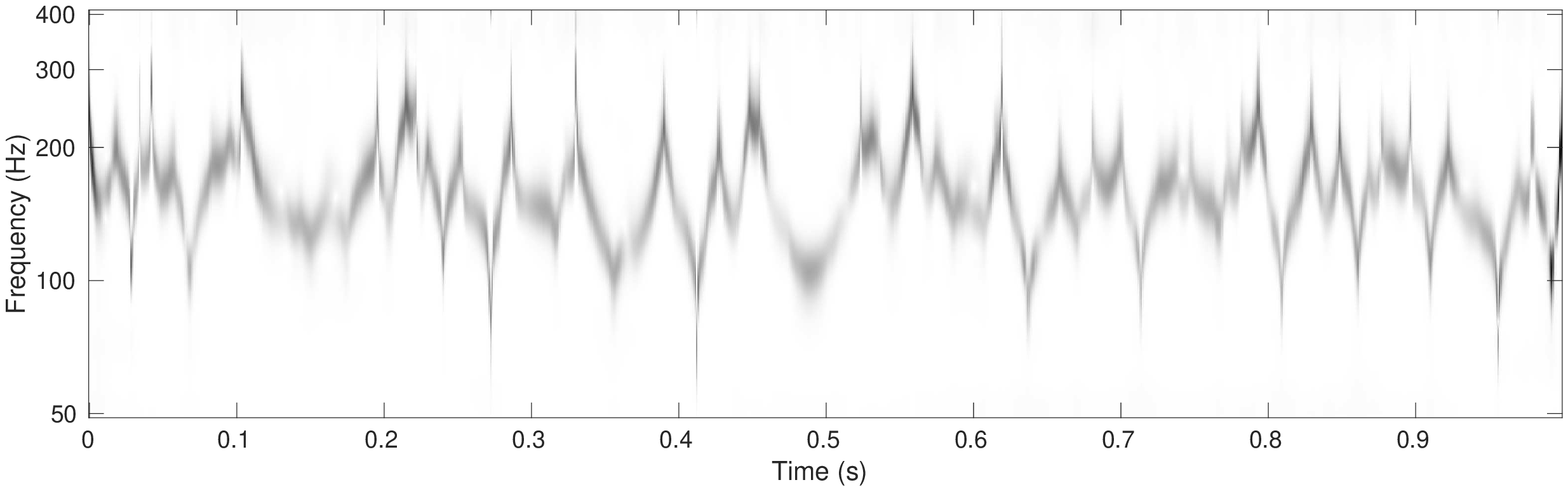}
\includegraphics[width=\textwidth]{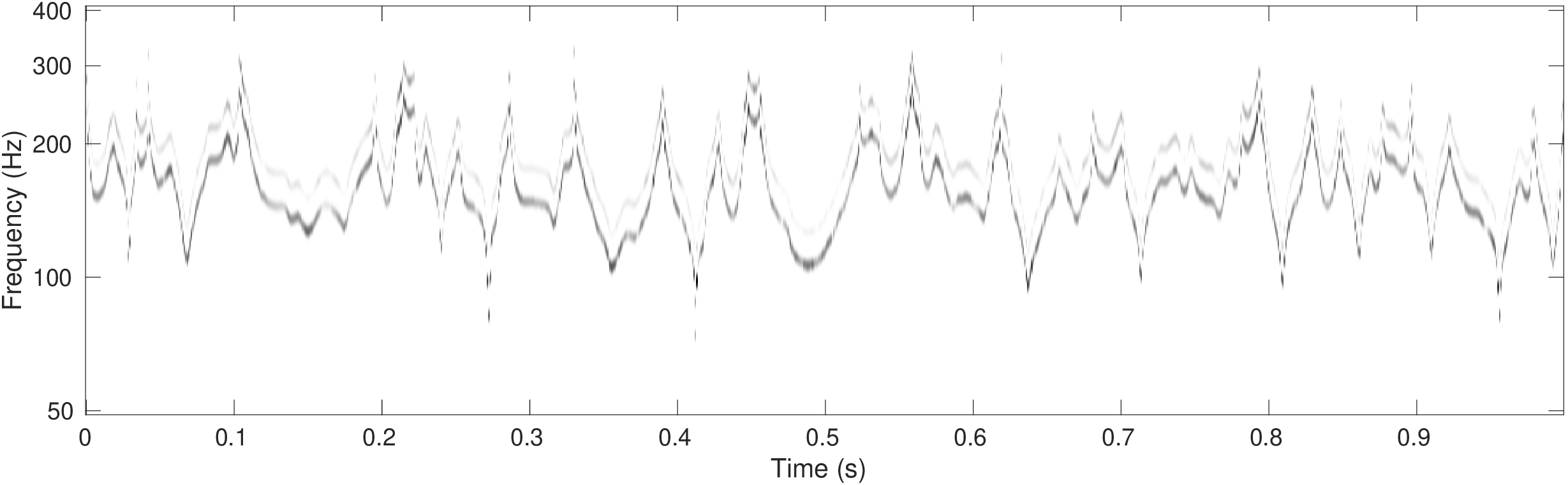}
\includegraphics[width=\textwidth]{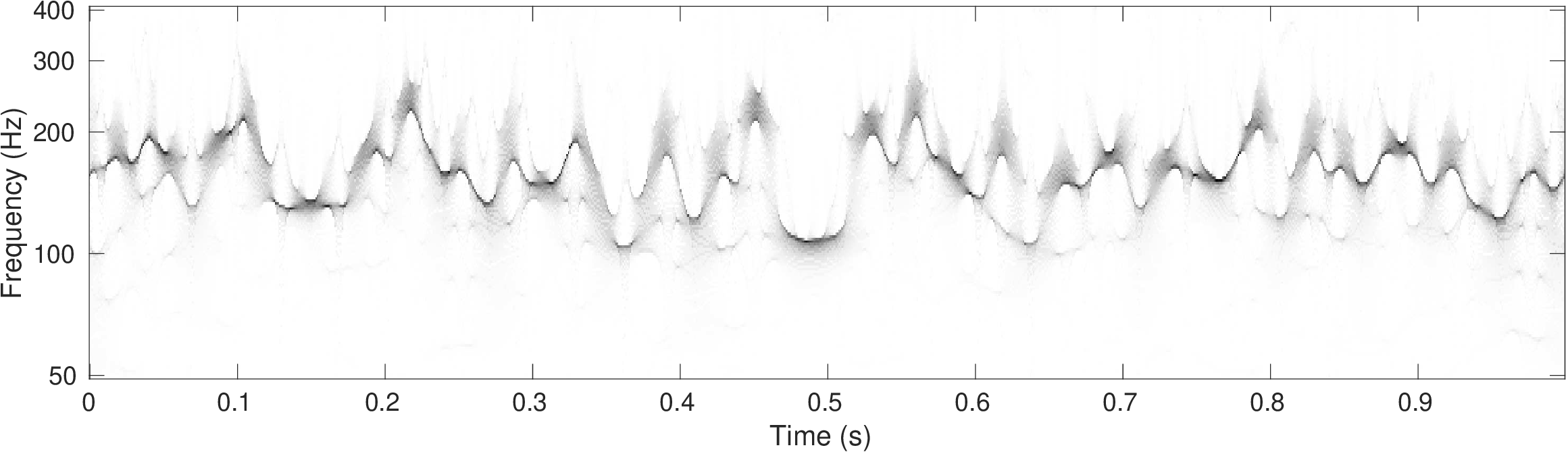}
\includegraphics[width=\textwidth]{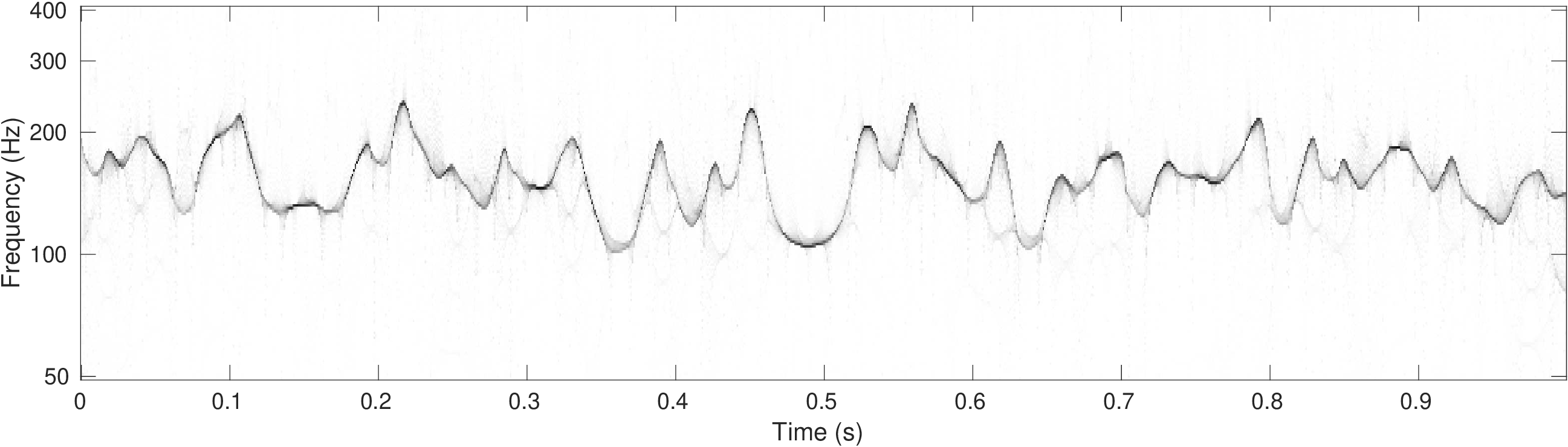}
\caption{JEFAS-S time-scale representation (first panel), and its sharpened version (second panel). The first (third panel) and second-order (fourth panel) synchrosqueezed wavelet transforms are below for comparison.}
\label{fi:comp.sst}
\end{figure}

Once the parameters of the model have been estimated, we can take advantage of the expression~\eqref{fo:JEFAS-S.WTestimation} to adapt the form of the time-scale representation to new a priori constraints. \edited{To do this, we simply change the expression of the covariance template $f$ to its sharp form provided by~\eqref{fo:sharp.spectrum}. Its implementation requires the estimates of the number of sine waves forming the stationary signal, their amplitudes and their frequencies. These estimates are obtained using the local maxima of the spectrum estimated by JEFAS-S. In this example, we find two local maxima, i.e $\tilde K = 2$. By construction, the wavelet-based spectrum estimator is smooth. The estimation of $K$ is therefore robust to noise. The estimated frequencies are $\tilde \xi_1 = 155$ and $\tilde\xi_2= 184$.} The resulting time-scale transform is shown in the middle Fig.~\ref{fi:comp.sst}, and as expected turns out to be much sharper than the time-scale representation given on the top image. In addition, the two components forming the signal appear quite distinctly, the synthesis was indeed able to separate these two components.

Besides, the comparison with the first-order synchrosqueezed wavelet transform, depicted in the third panel of Fig.~\ref{fi:comp.sst}, shows that our sharpened representation presents a much better concentration of the time-scale coefficients along the instantaneous frequencies, and does not suffer from the smoothing effect which is visible on the synchrosqueezed scalogram. \edited{The second-order synchrosqueezed wavelet transform~\cite{Oberlin2017second},  displayed in the fourth panel of Fig.~\ref{fi:comp.sst}, accounts for the rapid variations of the instantaneous frequency to allow for better localization in the time-frequency plane, but does not dissociate the two frequency components either. In addition, it tends to oversmooth the time warping function in comparison with JEFAS-S.

State-of-the-art synchrosqueezed transforms are able to provide sharp representations for signals containing multiple components with unnecessarily related instantaneous frequencies (see e.g.~\cite{Li2020Adaptive}). In particular, methods that handle crossover instantaneous frequency curves are now available~\cite{Chui2021Time}. This source separation task is not tractable with our method, as the model assumes simultaneous evolution of the instantaneous frequencies. Our approach, as a by-product of JEFAS-S, is helpful when a single non-stationarity source is present.
}

\section{Conclusions}
\label{se:conclu}
We have developed in this paper a new approach for time-scale representation of a class of non-stationary signals, which is able to handle fast varying non-stationarities. In comparison with existing approaches, a main originality is that JEFAS-S is based upon explicit probabilistic modeling (in the spirit of~\cite{Turner2014time}) of a time-scale synthesis representation, which allows to bypass (to a certain extent) constraints due to uncertainty principles.

JEFAS-S can handle both locally broadband and narrowband signals. In the latter case, i.e. for signals generated by time warping of sums of sine waves, JEFAS-S is also able to yield extremely sharp time-scale representations, sharper and much less biased than reassigned or synchroszqueezed spectrograms. Since JEFAS-S is a synthesis based approach, we also point out that the signal can be synthesized from the sharpened representation (as well as a stationary signal constructed from a corrected time-scale transform).

The approach is limited so far to non-stationarities generated by Time Warping (TW), but could be extended to other types, for example by adding amplitude modulation, or introducing frequency modulation or more general frequency nonlinear transformation (possibly using a more general time-frequency representation within which the nonlinearity could be described by a translation on a nonlinearly modified frequency axis).


\bibliography{TimeFreq}
\appendix

\section{Complex Gaussian distributions}
\label{app:complex.gaussian}
We recall here basic facts on multivariate complex Gaussian distributions, referring to~\cite{Picinbono1996second} for more details.

\begin{enumerate}
\item
A complex random vector $\bW\in\CC^N$ follows a complex Gaussian distribution if the real random vector 
\[
\bZ=\begin{pmatrix}\bX\\\bY
\end{pmatrix}=\begin{pmatrix}\Re(\bW)\\\Im(\bW)\end{pmatrix}\in\RR^{2N}
\]
follows a real Gaussian distribution $\cN\left(\bmu_\bZ,\bGamma_\bZ\right)$.
\item
A complex Gaussian distribution is characterized by its mean $\bmu\in\RR^N$, and its covariance and relation matrices $\bGamma\in\CC^{N\times N}$ and $\bR\in\CC^{N\times N}$ defined by
\begin{equation}
\bmu = \Ex{\bW}\ ,\qquad \bGamma = \Ex{\bW\bW^H} - \bmu\bmu^H\ ,\qquad \bR = \Ex{\bW\bW^T} - \bmu\bmu^T\ ,
\end{equation}
the superscript $H$ denoting Hermitian conjugation and the superscript $T$ denoting transposition. $\bGamma$ is then Hermitian symmetric, and $\bR$ is symmetric.
\item
The mean and covariance matrix of the associated real random vector $\bZ$ are given by
\begin{equation}
\bmu_\bZ = \begin{pmatrix}\Re(\bmu)\\\Im(\bmu)\end{pmatrix}\ ,\qquad
\bGamma_\bZ = \frac{1}{2}\begin{pmatrix}
\Re(\bGamma+\bR)&\Im(-\bGamma+\bR)\\\Im(\bGamma+\bR)&\Re(\bGamma-\bR)
\end{pmatrix}\ .
\end{equation}
\item
The distribution of a complex Gaussian random vector with mean $\bmu$, covariance matrix $\bGamma$ and relation matrix $\bR$ is denoted by $\cCN(\bmu,\bGamma,\bR)$, and takes the form
\begin{equation}
\label{fo:complex.gaussian.pdf}
p(\bZ) = K \exp\left\{
-\frac{1}{2}
\begin{pmatrix}\overline{\bZ}-\overline{\bmu}\\ \bZ-\bmu\end{pmatrix}^T
\begin{pmatrix}\bGamma & \bR\\\overline{\bR}&\overline{\bGamma}\end{pmatrix}^{-1}
\begin{pmatrix}\bZ-\bmu\\\overline{\bZ}-\overline{\bmu}\end{pmatrix}
\right\}\ ,
\end{equation}
for some constant $K$. The inverse of the above block covariance-relation matrix takes the form
\begin{equation}
\label{fo:inv.cov.rel}
\begin{pmatrix}\bGamma & \bR\\\overline{\bR}&\overline{\bGamma}\end{pmatrix}^{-1}
\!\!\!\!=\begin{pmatrix}\bA&\bB\\\overline{\bB}&\overline{\bA}\end{pmatrix}\ ,\quad\text{with}\quad
\left\{
\begin{array}{lll}
\bA \!\!&\!\!=\!\!&\!\! \left(\bGamma - \bR\overline{\bGamma}^{-1}\overline{\bR}\right)^{-1} \\
\bB \!\!&\!\!=\!\!&\!\! -\bA \bR\overline{\bGamma}^{-1}
\end{array}
\right. \!\! .
\end{equation}

\item
The complex Gaussian random vector $\bW$ is called circularly symmetric, or circular, if for every (deterministic) $\varphi\in (-\pi,\pi]$, $e^{i\varphi}\bW$ and $\bW$ have the same distribution. In such a case, $\bmu=0$ and $\bR=0$, one then writes $\bW\sim\cCN\!_c(0,\bGamma)$.
\end{enumerate}

\section{Proof of Proposition~\ref{prop:posterior}}
\label{app:prop1.proof}
Being the composition of complex Gaussian distributions, the posterior distribution of $\bW$ given $\by$ is complex Gaussian, non-symmetric, and its probability density function takes the form
\begin{equation}
\bW\vert\by \sim \cCN(\bmu_\mathsf{po},\bGamma_\mathsf{po},\bR_\mathsf{po})
\end{equation}
Therefore, using the notations of~\eqref{fo:complex.gaussian.pdf} for the inverse covariance-relation matrix, with $\bmu=\bmu_\mathsf{po}$, $\bGamma = \bGamma_\mathsf{po}$ and $\bR=\bR_\mathsf{po}$, we obtain
\begin{equation}
\nonumber
\ln p(\bW\vert\by) = K' -\bW^H\bA\bW -\Re\left(\bW^T\overline{\bB}\bW\right)
+2\Re\left(\bW^H\left(\bA\bmu_\mathsf{po} + \bB\overline{\bmu}_\mathsf{po}\right)\right)\ .
\end{equation}
Using~\eqref{fo:W.prior} and~\eqref{fo:obs.likelihood} together with Bayes' formula yields
\begin{eqnarray*}
\ln p(\bW\vert\by) &=& K' -\frac{1}{2\sigma^2}\left(\left(\by-\Re\left(\bD\bW\right)\right)^T \left(\by-\Re\left(\bD\bW\right)\right)\right)-\bW^H\bGamma_\mathsf{pr}^{-1}\bW\\
&=& K'' - \frac{1}{4\sigma^2} \bW^H\left(\bD^H\bD + 4\sigma^2\bGamma_\mathsf{pr}^{-1}\right)\bW \\ 
&&\hphantom{K" aaaaa}-\frac{1}{4\sigma^2}\Re\left(\bW^T (\bD^T\bD)\bW\right)  +\frac{1}{\sigma^2}\Re\left(\bW^H\bD^H\bY\right)\ .
\end{eqnarray*}
By identification with the previous expression, we obtain the parameters
\begin{equation*}
\bA = \frac{1}{4\sigma^2}\bD^H\bD +\bGamma_\mathsf{pr}^{-1}\ ,\qquad
\bB = \frac{1}{4\sigma^2}\overline{\bD}^T\overline{\bD}\ ,\quad
\bA\bmu_\mathsf{po} + \bB\overline{\bmu}_\mathsf{po} = \frac{1}{2\sigma^2}\bD^H\by\ .
\end{equation*}
After some algebra, we obtain that $\bmu_\mathsf{po}$ is a solution of the equation
\[
\bmu_\mathsf{po} = \frac{1}{2\sigma^2}\bGamma_\mathsf{pr}\bD^H\left(\by -\Re\left(\bD\bmu_\mathsf{po}\right)\right)\ .
\]
Setting $\bz = \Re(\bD\mu_\mathsf{po})$, then $2\sigma^2\bz = \Re\left(\bD\bGamma_\mathsf{pr}\bD^H\right)(\by-\bz)$ which yields
\[
\bz = \left( 2\sigma^2\Id + \Re(\bD\bGamma_\mathsf{pr}\bD^H)\right)^{-1} \Re\left(\bD\bGamma_\mathsf{pr}\bD^H\right)\by
\]
and therefore
\[
\bmu_\mathsf{po} = \frac{1}{2}\bGamma_\mathsf{pr}\bD^H\bC_\by^{-1}\by\ ,
\]
where
\begin{eqnarray*}
\bC_\by &=& \sigma^2\left(\Id - \left( 2\sigma^2\Id + \Re(\bD\bGamma_\mathsf{pr}\bD^H)\right)^{-1}\Re\left(\bD\bGamma_\mathsf{pr}\bD^H\right)
\right)^{-1}\\
&=& \sigma^2\Id + \frac{1}{2}\Re\left(\bD\bGamma_\mathsf{pr}\bD^H\right)
\end{eqnarray*}
the latter expression following from Woodbury's formula, which states that for every conformable matrices $A,U,C,V$, one has $(A+UCV)^{-1} = A\inv - A\inv U(C\inv + VA\inv U)\inv VA\inv$. This proves~\eqref{fo:mu.po}.

\medskip
We now turn to the expression of the posterior covariance matrix. Equating the expressions in~\eqref{fo:inv.cov.rel} with the expressions obtained above yields
\begin{eqnarray*}
\bGammapo - \bR_\mathsf{po}\obGammapo^{-1}\overline{\bR}_\mathsf{po}&=&4\sigma^2\left(\bD^H\bD+4\sigma^2\bGammapr^{-1}\right)^{-1}\\
\bR_\mathsf{po} &=& -\left(\bD^H\bD + 4\sigma^2\bGammapr^{-1}\right)^{-1}\obD^T\obD\,\obGammapo\ .
\end{eqnarray*}
Inserting the expressions of $\bR_\mathsf{po}\obGammapo^{-1}$ and $\bR_\mathsf{po}^{-1}$ into the first of the two above equations yields
\begin{eqnarray*}
\bGammapo \!\!&\!\!=\!\!&\!\! \left[4\sigma^2\,\Id + \bD^H\bD - \obD^T\obD\left(\obD^H\obD\!+\!4\sigma^2\obGammapr^{-1}\right)^{-1}\!\!\bD^T\bD \right]^{-1}\!\!\!\left(\bD^H\bD \!+\! 4\sigma^2\bGammapr^{-1}\right)^{-1}\\
\!\!&\!\!=\!\!&\!\! \left[\bGammapr^{-1} +\frac{1}{4\sigma^2}\bD^H\left[\Id-\obD\left(\obD^H\obD + 4\sigma^2\obGammapr^{-1}\right)^{-1}\bD^T\right]\bD\right]^{-1}\ .
\end{eqnarray*}
Observe that by Woodbury's formula we have that
\begin{eqnarray*}
\left[\Id-\obD\left(\obD^H\obD + 4\sigma^2\obGammapr^{-1}\right)^{-1}\bD^T\right]^{-1}
\!\!\!\!&\!\!=\!\!&\Id -\obD\left(\bD^T\obD - \left(\obD^H\obD + 4\sigma^2\obGammapr^{-1}\right)\right)\bD^T\\
\!\!&\!\!=\!\!& \Id +4\sigma^2\obD\,\obGammapr^{-1}\bD^T
\end{eqnarray*}
so that, setting $\bC=(4\sigma^2)\inv \left(\Id +\obD\left(\obD^H\obD + 4\sigma^2\obGammapr^{-1}\right)^{-1}\bD^T\right)$
\[
\bGammapo =\left[\bGammapr^{-1} +\bD^H\bC\bD\right]^{-1}= 
\bGammapr - \bGammapr\bD^H\left(\bC^{-1} + \bD\bGammapr\bD^H\right)^{-1}\bD\bGammapr\ .
\]
Now
\[
\bC^{-1} = 4\sigma^2 \left[\Id - \obD\left(-\obD^H\obD - 4\sigma^2\obGammapr^{-1} + \bD^T\obD\right)^{-1}\bD^T\right]=4\sigma^2\Id +\obD\,\obGammapr\bD^T\ ,
\]
therefore we obtain
\[
\bGammapo = \bGammapr - \bGammapr\bD^H\left(4\sigma^2\Id + \bD\bGammapr\bD^H + \obD\,\obGammapr\bD^T\right)^{-1}\bD\bGammapr
\]
which yields~\eqref{fo:Gamma.po}.\hfill$\square$

\section{Proof of Proposition~\ref{prop:EM.update}}
\label{app:proof.EM.update}

We first notice that according to~\eqref{fo:JEFAS-S.WTestimation}, the estimate for the time-scale representation given $\bvtheta^{(k-1)}$ reads
\begin{equation}
\tbw_n^{(k)} = \frac{1}{2} \tbC_n^{(k-1)}\bPsi_n^H\left(\tbC_\by^{(k-1)}\right)^{-1}\by\ ,
\end{equation}
which is the desired expression.
EM proceeds by alternating the expectation and maximization steps, which are described below.
\subsection{Expectation}
The goal is to evaluate the quantity defined in~\eqref{fo:EM.Q}, which may be written as
\begin{equation}
Q\left(\bvtheta,\tbvtheta^{(k-1)}\right)= \EE_{(k-1)}\!\left\{\ln\left(p(\by\vert\bW)\right)\right\} + \EE_{(k-1)}\!\left\{\ln\left(p_{\bvtheta}(\bW)\right)\right\}\ .
\end{equation}
According to~\eqref{fo:obs.likelihood}, the first term does not depend on $\bvtheta$ and can be ignored in the optimization. The second term reads
\begin{eqnarray*}
\EE_{(k-1)}\!\left\{\ln\left(p_{\bvtheta}(\bW)\right)\right\}&=&-\ln\det(\bGammapr(\bvtheta))-N_\tau M_s\ln(\pi)\\&&\hphantom{aaaaaaaa}-\EE_{(k-1)}\!\left\{\bW^H\bGammapr(\bvtheta)^{-1}\bW\right\}\ .
\end{eqnarray*}
Focusing on the last term, setting $\bX=\Re(\bW)$, $\bY=\Im(\bW)$ and $\bZ=(\bX^T,\bY^T)^T$, we have
\begin{eqnarray*}
\EE_{(k-1)}\!\left\{\!\bW^H\!\bGammapr(\bvtheta)^{-1}\bW\!\right\}\!\!&\!\!\!=\!\!&\!\! \EE_{(k-1)}\!\left\{\!\bX^T\!\bGammapr(\bvtheta)^{-1}\bX\!\right\}\!+\!\EE_{(k-1)}\!\left\{\!\bY^T\!\bGammapr(\bvtheta)^{-1}\bY\!\right\}\\
\!\!&\!\!\!=\!\!&\!\!\EE_{(k-1)}\!\left\{\!\bZ^T\!
\begin{pmatrix}\bGammapr(\bvtheta)^{-1}&0\\0&\bGammapr(\bvtheta)^{-1}\end{pmatrix}\bZ\!\right\}\ .
\end{eqnarray*}
Now, since at iteration $k$ and conditionally to $\by$, $\bZ\vert\by\sim\cCN\left(\tbm^{(k-1)},\tbG^{(k-1)}\right)$, with
\begin{eqnarray*}
\tbm^{(k-1)} &=& \begin{pmatrix}
\Re\left(\tbmu_\mathsf{po}^{(k-1)}\right)\\
\Im\left(\tbmu_\mathsf{po}^{(k-1)}\right)
\end{pmatrix}\ ,\\
\tbG^{(k-1)} &=& \frac{1}{2}\!\begin{pmatrix}
\Re\!\left(\tbGammapo^{(k-1)}\!+\! \tbRpo^{(k-1)}\right)&
\Im\!\left(-\tbGammapo^{(k-1)}\!+\! \tbRpo^{(k-1)}\right)\\
\Im\!\left(\tbGammapo^{(k-1)}\!+\! \tbRpo^{(k-1)}\right)&
\Re\!\left(\tbGammapo^{(k-1)}\!-\! \tbRpo^{(k-1)}\right)
\end{pmatrix}\ ,
\end{eqnarray*}
and where we have set for simplicity 
\[
\tbmu_\mathsf{po}^{(k-1)}=\bmu_\mathsf{po}\!\left(\tbvtheta^{(k-1)}\!\right)\,,\ \tbGammapo^{(k-1)} = \bGammapo\!\left(\tbvtheta^{(k-1)}\!\right)\,,\ 
\tbRpo^{(k-1)} = \bRpo\!\left(\tbvtheta^{(k-1)}\!\right)\ .
\]
Therefore we have that
\begin{eqnarray*}
\EE_{(k-1)}\!\left\{\!\bW^H\!\bGammapr(\bvtheta)^{-1}\bW\!\right\}\!\!&\!\!\!=\!\!&\!\!\Tr{\begin{pmatrix}
\bGammapr(\bvtheta)^{-1}&0\\0&\bGammapr(\bvtheta)^{-1}\end{pmatrix}\tbG^{(k-1)}}\\
&&+ \begin{pmatrix}
\Re\left(\tbmupo^{(k-1)}\right)\\\Im\left(\tbmupo^{(k-1)}\right)
\end{pmatrix}^T\bGammapr(\bvtheta)^{-1}\begin{pmatrix}
\Re\left(\tbmupo^{(k-1)}\right)\\\Im\left(\tbmupo^{(k-1)}\right)
\end{pmatrix}\\
&=&\Tr{\bGammapr(\bvtheta)^{-1}\Re\left(\tbGammapo^{(k-1)}\right)}\\
&&\hphantom{aaaaaa} + \left(\tbmupo^{(k-1)}\right)^H \bGammapr(\bvtheta)^{-1}\tbmupo^{(k-1)}\ .
\end{eqnarray*}
Putting things together we finally obtain
\begin{eqnarray}
\nonumber
Q\left(\bvtheta,\tbvtheta^{(k-1)}\right) &=& -\ln\det(\bGammapr(\bvtheta)) -\Tr{\bGammapr(\bvtheta)^{-1}\Re\left(\tbGammapo^{(k-1)}\right)}\\
&&\hphantom{aaaaaa}- \left(\tbmupo^{(k-1)}\right)^H \bGammapr(\bvtheta)^{-1}\tbmupo^{(k-1)} + C\ ,
\label{fo:Q}
\end{eqnarray}
for some additive constant $C$.

\subsection{Maximization}

Plugging the expressions of $\bGammapr(\bvtheta)$ and $\bGammapo^{(k-1)}$ into~\eqref{fo:Q}, we readily obtain
\begin{eqnarray*}
\tbvtheta^{(k)} &=& \mathop{\mathsf{arg\,min}}_{\bvtheta}
\sum_{n=1}^{N_\tau}\bigg[ \ln\det(\bC(\vartheta_n)) + \tbw_n^{(k-1)\,H}\bC(\vartheta_n)^{-1}\tbw_n^{(k-1)}\\
&&\hphantom{aaaaaaaaaaaaaaaaaaaaa}+\Tr{\bC(\vartheta_n)^{-1}\Re\left(\tbGamma_n^{(k-1)}\right)}\bigg]
\end{eqnarray*}
which is the desired expression.


\end{document}